\documentclass[oneside,twocolumn,aps,prl,preprintnumbers,bibnotes10pt,superscriptaddress,amsmath,amssymb]{revtex4-2}
\usepackage{graphicx,subfigure,wrapfig}
\usepackage{float}
\usepackage{color}
\usepackage{epstopdf}
\usepackage{amsmath}
\usepackage{braket}
\usepackage{amsthm}
\usepackage{amssymb}
\usepackage{amsfonts}
\usepackage{graphicx,subfigure}
\usepackage{txfonts}
\usepackage{ulem}
\usepackage{mathtools}
\usepackage{bm}
\usepackage{bbm}
\usepackage{hyperref}
\usepackage{natbib}
\usepackage{comment}

\newcommand{\Eins}{\ensuremath{\mathbbm 1}}

\newcommand{\be}{\begin{equation}}
\newcommand{\ee}{\end{equation}}
\newcommand{\beq}{\begin{eqnarray}}
\newcommand{\eeq}{\end{eqnarray}}
\newcommand{\vect}[1]{\bm{#1}}
\newcommand{\makeref}[1]{(\ref{#1})}

\begin{document}

\title{Distributed quantum multiparameter estimation with optimal local measurements} 

\author{L. Pezzè}
\affiliation{Istituto Nazionale di Ottica del Consiglio Nazionale delle Ricerche (CNR-INO), Largo Enrico Fermi 5, 50125 Firenze, Italy}
\affiliation{European Laboratory for Nonlinear Spectroscopy, Università degli studi di Firenze, Via N. Carrara 1, 50019 Sesto Fiorentino, Italy}
\affiliation{QSTAR, Largo Enrico Fermi 2, 50125 Firenze, Italy}

\author{A. Smerzi}
\affiliation{Istituto Nazionale di Ottica del Consiglio Nazionale delle Ricerche (CNR-INO), Largo Enrico Fermi 5, 50125 Firenze, Italy}
\affiliation{European Laboratory for Nonlinear Spectroscopy, Università degli studi di Firenze, Via N. Carrara 1, 50019 Sesto Fiorentino, Italy}
\affiliation{QSTAR, Largo Enrico Fermi 2, 50125 Firenze, Italy}

\begin{abstract}
We study the multiparameter sensitivity bounds of a sensor made by an array of $d$ spatially-distributed Mach-Zehnder interferometers (MZIs). 
A generic single non-classical state is mixed with $d-1$ vacuums to create a $d$-modes entangled state, each mode entering one input port of a MZI, while a coherent state enters its second port. 
We show that local measurements, independently performed on each MZI, are sufficient to provide a sensitivity saturating the quantum Cram\'er-Rao bound. 
The sensor can overcome the shot noise limit for the estimation of arbitrary linear combinations of the $d$ phase shifts, provided that the non-classical probe state has an anti-squeezed quadrature variance.
We compare the sensitivity bounds of this sensor with that achievable with $d$ independent MZIs, each probed with a nonclassical state and a coherent state. 
We find that the $d$ independent interferometers can achieve the same sensitivity of the entangled protocol but at the cost of using additional $d$ non-classical states rather than a single one. 
When using in the two protocols the same average number of particles per shot $\bar{n}_T$, we find analytically a sensitivity scaling $1/\bar{n}_T^2$ for the entangled case which provides
a gain factor $d$ with respect to the separable case where the sensitivity scales as 
$d/\bar{n}_T^2$.
We have numerical evidences that the gain factor $d$ is also obtained when fixing the total average number of particles, namely when optimizing with respect to the number of repeated measurements.
\end{abstract}

\maketitle

\begin{figure}[t!]
\includegraphics[width=1\columnwidth]{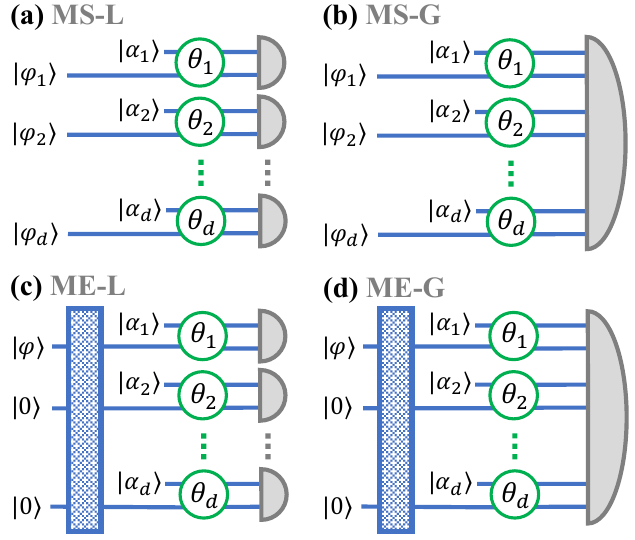}
\caption{Distributed quantum multiparameter estimation can follow four different schemes depending on the presence/absence of mode-entanglement in the probe state and performing local/global output measurements.
Each green circle represents schematically a Mach-Zehnder interferometer with two input and two output modes.
Blue shaded box is a linear lossless devices (quantum circuit) generating mode entanglement.
The gray semicircle indicate detection.
We distinguish four schemes: 
(a) mode-separable probe and local measurements (MS-L);
(b) mode-separable probe and global measurements (MS-G);
(c) mode-entangled probe and local measurements (ME-L); and
(d) mode-entangled probe and global measurements (ME-G).} \label{Figure1}
\end{figure}

{\it Introduction.---}Distributed multiparameter estimation aims to simultaneously infer parameters encoded in spatially-allocated quantum devices. 
This is a central task in the emerging technology of quantum sensing~\cite{YePRL2024}, with applications in sensor networks~\cite{KomarNATPHYS2014, PolzikPRA2016, ProctorPRL2018, GessnerNATCOMM2020, BradyPRXQUANTUM2022}, biological probing~\cite{TaylorPHYSREP2016, CasacioNATURE2021}, imaging~\cite{RehacekPRA2017,MoreauNRP2019}, inertial~\cite{GracePRAPP2020,GoldbergJP2021} and magnetic field~\cite{BaumgratzPRL2016, ApellanizPRA2018, HouPRL2020, LipkaPRApp2024} sensing.
Computing the ultimate sensitivity bounds to guide the complete optimization of the estimation protocol is a complex and multifaceted problem, encompassing the selection of probe states, allocation of resources, choice of measurement observables, and formulation of estimation strategies~\cite{AlbarelliPLA2020, DemkowiczJPA2020}.
Four schemes are possible, contingent on the utilization of mode-separable (MS) or mode-entangled (ME) probe states and the implementation of local (L) or global (G) measurements: these are indicated as MS-L, MS-G, ME-L and ME-G protocols in Fig.~\ref{Figure1}.
Mode entanglement can be created using a linear lossless device that extends the conventional two-mode beam-splitter to the case of $d$ modes~\cite{ReckPRL1994, GuoNATPHYS2020, XiaPRL2020, GePRL2018}.
This is indicated as quantum circuit (QC) and is schematically represented in Fig.~\ref{Figure1} by a blue box.
A local measurement entails detecting the output modes of each sensor individually. 
While certain multiparameter schemes employing a single common reference mode inherently necessitate global measurements~\cite{HumphreysPRL2013, CiampiniSCIREP2016, ValeriNPJ2020, OhPRR2020, GagatsosPRA2016, LiNJP2020, GoldbergPRA2020, HongNATCOMM2021}, local measurements circumvent the need to recombine all the modes to realize a global POVM~\cite{GuoNATPHYS2020, XiaPRL2020, MaliaNATURE2022, ZhuangPRA2018, LiuNATPHOT2021}.

In this manuscript, we consider distributed multiparameter estimation with local measurements, as in Fig.~\ref{Figure1}(a) and~(c).
In this case, the most general multiparameter sensor is provided by an array of $d$ Mach-Zehnder interferometers (MZIs), where each MZI sense the relative phase shift among two interferometric arms (schematically represented in Fig.~\ref{Figure1} by a green circle). 
We estimate each of the $d$ unknown parameters $\vect{\theta} = \{ \theta_1, ..., \theta_d\}$, $\theta_j$ being the relative phase in the $j$th MZI.
The sensors configuration is optimized with the goal to minimize $\Delta^2(\vect{v} \cdot \vect{\theta})$, where $\vect{v} = \{v_1, ..., v_d\}$ is an arbitrary vector of real coefficients~\cite{XiaPRL2020, OhPRR2020, RubioJPA2020, GattoPRR2021, MalitestaPRA2023, GrossJPA2021, QianPRA2019, BringewattPRR2024, nota1} with normalization $\sum_{j=1}^d \vert v_j \vert=1$.

The four different schemes of Fig.~\ref{Figure1}(a)-(d) involve $2d$ modes as input.
These consist of $d$ coherent states $\ket{\alpha_1} \otimes ...\otimes \ket{\alpha_d}$, with average number of particles $\bar{n}_c = \sum_{j=1}^d \vert \alpha_j \vert^2$, and $d$ generic single-mode states $\ket{\varphi_1}\otimes...\otimes \ket{\varphi_d}$, with average number of particles $\bar{n} = \sum_{j=1}^d  \bar{n}_j$, where $\bar{n}_j$ is the intensity of $\ket{\varphi_j}$.
The protocols~(a) and~(b) of Fig.~\ref{Figure1} use MS probe states.
In this case the $j$th MZI is fed by $\ket{\alpha_j}$ in one input and by $\ket{\varphi_j}$ on the other input.
We first consider the case $\ket{\varphi_j} = \ket{\varphi}$ for all $j$.
The protocols~(c) and~(d) involve a ME probe state, with the QC mixing the $d$ states $\ket{\varphi}\otimes \ket{0} \otimes ... \otimes \ket{0}$, where $\ket{0}$ is the vacuum.
In this case, the $j$th MZI is fed by $\ket{\alpha_j}$ in one input, while the other input is the $j$th output mode of the QC.
Let us indicate with $\Delta^2 (\vect{v} \cdot \vect{\theta})$ the quantum Cram\'er-Rao bound (QCRB) sensitivity optimized over the intensities of the $d$ coherent states and (in the case of a ME probe states) over arbitrary QC operations.
Remarkably, such an optimization can be performed analytically for all schemes~(a)-(d) of Fig.~\ref{Figure1}, giving~\cite{supp}
\be \label{Eq.Sopt}
\Delta^2 (\vect{v} \cdot \vect{\theta}) 
= \frac{1}{m[ \bar{n}_c \, (\Delta^2 \hat{p})_{\ket{\varphi}} + \bar{n}]}.
\ee
Here, $m$ is the number of measurements ($m\gg 1$ is assumed to for the saturation of the QCRB) and $(\Delta^2 \hat{p})_{\ket{\varphi}}$ is the single-mode $p$-quadrature variance of the state $\ket{\varphi}$.
Clearly, quantum states with large $(\Delta^2 \hat{p})_{\ket{\varphi}}$ (such as Gaussian squeezed, Schr\"odinger cat or Fock states) are desired.
If we further optimize $\Delta^2 (\vect{v}\cdot\vect{\theta})$ for the MS schemes over the states $\ket{\varphi_1} ... \ket{\varphi_d}$ -- considered here to be not necessarily equal -- we obtain~\cite{supp}
\be \label{Eq.SHL}
\Delta^2 (\vect{v}\cdot\vect{\theta})_{\rm MS}  = \min_{\ket{\varphi_1},...,\ket{\varphi_d}} \Delta^2 (\vect{v}\cdot\vect{\theta}) 
= 
\frac{\lVert \vect{v} \rVert_{2/3}^2}{m\bar{n}_T^2},
\ee
where $\lVert \vect{v} \rVert_\kappa = (\sum_{j=1}^d \vert v_j \vert^\kappa)^{1/\kappa}$ is the $\kappa$-norm of the vector $\vect{v}$.
Instead, when optimizing $\Delta^2 (\vect{v}\cdot\vect{\theta})$ for the ME schemes over the state $\ket{\varphi}$, we find 
 \be \label{Eq.EHL}
\Delta^2 (\vect{v}\cdot\vect{\theta})_{\rm ME}  = \min_{\ket{\varphi}} \Delta^2 (\vect{v} \cdot \vect{\theta}) 
=
\frac{1}{m\bar{n}_T^2}.
\ee
In both Eqs.~(\ref{Eq.SHL}) and (\ref{Eq.EHL}), $\bar{n}_T = \bar{n}_c + \bar{n}$ is the total average number of particles per measurement shot and the minimum is achieved by squeezed-vacuum state(s), as discussed below.
The $\vect{v}$-dependent prefactor in Eqs.~(\ref{Eq.Sopt})-(\ref{Eq.EHL}) takes into account the optimal distribution of resources for the given combination $\vect{v}$ of parameters.
Interestingly, we show that the optimized QCRBs Eqs.~(\ref{Eq.Sopt})-(\ref{Eq.EHL}) can be saturated without global measurements. 
In particular, number-counting at the output ports of each MZI is optimal, in the sense that it allows to saturate the QCRB with a $\vect{\theta}$-independent sensitivity. 
Entangled measures, as in the MS-G scheme of Fig.~\ref{Figure1}(b) or the ME-G scheme of Fig.~\ref{Figure1}(d), are not necessary to saturate the QCRB, in our case.

First, let us notice that Eqs.~(\ref{Eq.Sopt})-(\ref{Eq.EHL}) recover known results for $d=1$~\cite{PezzePRL2008, LangPRL2013, PezzePRL2013}. 
Furthermore, when $(\Delta^2 \hat{p})_{\ket{\varphi}}=1$ (namely for the vacuum $\ket{\varphi}=\ket{0}$ or the coherent $\ket{\varphi}=\ket{\alpha}$ states), the MS-L and the ME-L schemes become equivalent when using the same  total average number of particles, $\bar{n}_T$.
In this case, Eq.~(\ref{Eq.Sopt}) predicts 
\be \label{Eq.SN}
\Delta^2 (\vect{v}\cdot\vect{\theta})_{\rm SN} =
\frac{1}{m\bar{n}_T},
\ee
corresponding to the shot-noise (SN) limit.
As shown by Eq.~(\ref{Eq.Sopt}), both MS-L and ME-L can overcome the SN provided that $(\Delta^2 \hat{p})_{\ket{\varphi}}>1$.
Notice that the condition $(\Delta^2 \hat{p})_{\ket{\varphi}}>1$ does not necessarily requires squeezing of the conjugate quadrature $q$, for instance in Fock states $\ket{\varphi}=\ket{n}$, where $(\Delta^2 \hat{p})_{\ket{n}} = (\Delta^2 \hat{q})_{\ket{n}} = 2n +1 $ are both anti-squeezed. 

One of the main consequences of our analysis is that, in the limit $\bar{n}_c \gg \bar{n}/(\Delta^2 \hat{p})\vert_{\ket{\varphi}}$ (Holstein-Primakoff approximation, where each MZI performs a quadrature displacement~\cite{XiaPRL2020, MalitestaPRA2023}), the ME-L scheme using a single state $\ket{\varphi}$ can perform equally well as the MS-L scheme, while the latter uses $d$ copies of the same state $\ket{\varphi}$: this represents a reduction of resource
overhead of high practical advantage.
If instead, we fix the total average number of particles per measurement shot, $\bar{n}_T$, and compare fully optimized MS-L and the ME-L schemes, we find the gain factor 
\be \label{G1}
\mathcal{G} = \frac{\Delta^2 (\vect{v} \cdot \vect{\theta})_{\rm MS}}{\Delta^2 (\vect{v} \cdot \vect{\theta})_{\rm ME}} = \lVert \vect{v} \rVert_{3/2}^2,
\ee
given by the ratio between Eq.~(\ref{Eq.SHL}) and Eq.~(\ref{Eq.EHL}).
Equation (\ref{G1}) ranges from 1 (for the estimation of a single parameter, corresponding to $v_k=1$ and $v_{j\neq k}=0$, such that $\lVert \vect{v} \rVert_{3/2}^2=1$) to $d$, when $v_j = \pm 1/d$ (such that $\lVert \vect{v} \rVert_{3/2}^2 = d$).
Finally, we point out that Eq.~(\ref{G1}) assumes the central limit ($m \gg 1$) in both schemes.
In principle, the number of repeated measurements needed to achieve the QCRB can be different for MS and ME strategies.
To address this issue, we have performed numerically a multiparameter maximum likelihood analysis.
When comparing MS-L and ME-L by optimizing the number of repeated measurements, we confirm a gain factor $d$, when $v_j = \pm 1$ for all $j$.
In the following, we discuss in more details the derivation and consequences Eqs.~(\ref{Eq.Sopt})-(\ref{G1}).

{\it Sensitivity bounds and optimal measurements.---}The QCRB for the four schemes of Fig.~\ref{Figure1}(a)-(d) is  $\Delta^2(\vect{v} \cdot \vect{\theta})_{\rm QCRB} = \vect{v}^T \vect{F}_{Q}^{-1} \vect{v}/m$, where $\vect{F}_Q$ is the $d\times d$ quantum Fisher information matrix (QFIM)~\cite{HelstromBOOK, AlbarelliPLA2020, DemkowiczJPA2020}:
$[\vect{F}_Q]_{(j,k)} = 4(\langle \Psi \vert \hat{H}_j \hat{H}_k \vert \Psi \rangle - \langle \Psi \vert \hat{H}_j \vert \Psi \rangle \langle \Psi \vert \hat{H}_k \vert \Psi \rangle)$ for the overall pure probe state $\ket{\Psi}$.
Here, $\hat{H}_j = (\hat{a}_j \hat{b}_j - \hat{a}_j \hat{b}_j)/(2i)$~\cite{YurkePRA1986}, $\hat{a}_j$ and $\hat{b}_j$ ($\hat{a}_j^\dag$ and $\hat{b}_j^\dag$) are bosonic annihilation (creation) operators for a particle in mode $a_j$ and $b_j$, respectively.
Spatial dislocation of the different MZIs guarantees that $[\hat{H}_j, \hat{H}_k]=0$, with $j,k=1,...,d$.
This commutativity condition implies that, for pure states, the QCRB can be achieved by optimal measurements~\cite{MatsumotoJPA2002, PezzePRL2017, YangPRA2019}. 
Number counting at the output ports of each MZI is described by the set of projectors $\{ \ket{\vect{\mu}}\bra{\vect{\mu}} \}_{\vect{\mu}}$, where $\ket{\vect{\mu}} = \ket{n_1}\otimes \ket{m_1} \otimes ... \otimes \ket{n_d}\otimes \ket{m_d}$,  $\ket{n_j}$ and $\ket{m_j}$ are Fock states for number of particles $n_j,m_j\geq 0$ at the two output ports $a_j$ and $b_j$, respectively, of the $j$th MZI. 
We demonstrate~\cite{supp} that number counting realizes a local measurement that fulfills the conditions for the saturation of the QCRB, provided that the coefficients $\bra{\vect{\mu}}\Psi\rangle$ are real.
When $\ket{\varphi}$ is the squeezed-vacuum state (see below), this condition is optimal, as demonstrated in Ref.~\cite{MalitestaPRA2023}.

\begin{figure*}[t!]
\includegraphics[width=1\textwidth]{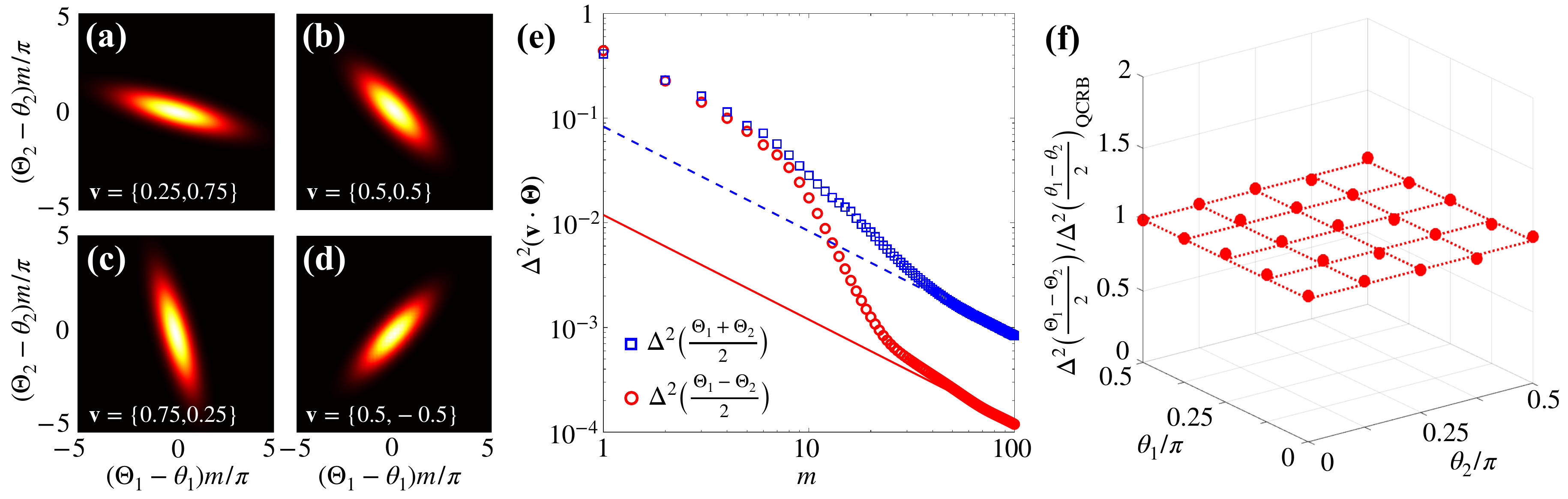}
\caption{Results of a maximum-likelihood analysis for the ME-L strategy with $d=2$.
Here, the sensor array is optimized according to Eq.~(\ref{Eq.optimal}), minimizing $\Delta^2(\vect{\theta} \cdot \vect{v})_{\rm QCRB}$ for a given $\vect{v}$.
(a)-(d) Maximum likelihood distributions, $P(\vect{\Theta}\vert \vect{\theta})$.
Each panel indicates the corresponding $\vect{v}$. 
$P(\vect{\Theta}\vert \vect{\theta})$ has its narrowest width along the $\vect{v}\cdot \vect{\theta
}$ axis. 
Here $m=100$ and $\vect{\theta}= \{\pi/2, \pi/2\}$.
(e) Maximum likelihood uncertainties $\Delta^2\big(\tfrac{\Theta_1-\Theta_2}{2}\big)$ (red circles) and $\Delta^2\big(\tfrac{\Theta_1+\Theta_2}{2}\big)$ (blue squares) as a function of $m$.
Here $\vect{\theta}= \{\pi/2, \pi/2\}$ and the sensor network is optimized for $\vect{v} = \{0.5,-0.5\}$.
The red solid line is Eq.~(\ref{Eq.Sopt}); the dashed blue line is the SN, Eq.~(\ref{Eq.SN}).
(f) Maximum likelihhood uncertainty $\Delta^2\big(\tfrac{\Theta_1-\Theta_2}{2}\big)$ (dots), normalized to the QCRB, Eq.~(\ref{Eq.Sopt}), and plotted as a function of $\theta_1$ and $\theta_2$, for $m=100$.
Dotted lines are guides to the eye. 
In all panels, we consider a Fock state as input $\ket{\varphi}=\ket{n}$ with $n = \bar{n}_T/2$ and $\bar{n}_T = 12$~\cite{supp}.
}
\label{Figure2}
\end{figure*}

{\it Optimized MS schemes.---}In the MS-L and MS-G schemes of Fig.~\ref{Figure1}(a) and (b), the overall probe state is $\ket{\Psi} = \bigotimes_{j=1}^d \ket{\psi_j}$, where $\ket{\psi_j}= \ket{\alpha_j} \otimes \ket{\varphi_j}$ is the input states of the $j$th MZI.
In this case the QFIM is diagonal, $[\vect{F}_Q]_{(j,k)} = \delta_{j,k} \vect{F}_Q^{(j)}$, where $F_Q^{(j)}$ is the quantum Fisher information of the $j$th sensor~\cite{supp, BraunsteinPRL1994}:
$F_Q^{(j)} = 4(\bra{\psi_j} \hat{H}_j^2 \ket{\psi_j} - \bra{\psi_j} \hat{H}_j \ket{\psi_j}^2)$ for pure states.
The QCRB reads $\Delta^2 (\vect{v}\cdot\vect{\theta})_{\rm QCRB} = \sum_{j=1}^d \vect{v}_j^2/(mF_Q^{(j)})$.
Equation~(\ref{Eq.Sopt}) is recovered by taking all $\ket{\varphi_j}$ equal to a generic $\ket{\varphi}$, and optimizing over the intensities $\vert \alpha_j \vert^2$ of each coherent state $\ket{\alpha_j}$, see \cite{supp}.
Equation~(\ref{Eq.SHL}) is obtained when further maximizing the intensity $\bar{n}_j$ and the quadrature variance $(\Delta^2 \hat{p})_{\ket{\varphi_j}}$ of each state $\ket{\varphi_j}$, giving $\vect{F}_Q^{(j)} = (\vert \alpha_j \vert^2 + \bar{n}_j)^2$~\cite{LangPRL2013}.
Further optimizing of all $\vert \alpha_j \vert^2$ and $\bar{n}_j$, for a given $\bar{n}_T$, provides Eq.~(\ref{Eq.SHL})~\cite{supp}.

{\it Optimized ME schemes.---}Mode entanglement created by the QC can result in off diagonal terms in the QFIM, in contrast to the case of a MS probe state. 
By an opportune choice of QC transformation and coherent state intensities, it is possible to engineer such off-diagonal terms in order to minimize $\Delta^2(\vect{v} \cdot \vect{\theta})_{\rm QCRB}$ for a given $\vect{v}$. To show this, we first notice that the QFIM can be written as $\vect{F}_Q = \gamma \vect{f} \vect{f}^T + \vect{D}$~\cite{supp}, where $\gamma=(\Delta^2 \hat{p})\vert_{\ket{\varphi}}-1$, $\vect{f}$ is a $d$-dimensional real vector with element $f_j = \vert \alpha_j \vert \sqrt{P_j}$, $P_j$ is the probability that a particle exits the output $b_j$ mode of the QC, and $\vect{D}$ is a diagonal $d\times d$ matrix with elements $D_{jj} = \vert \alpha_j\vert^2 + P_j \bar{n}$.
Remarkably the $2d\times 2d$ QFIM can be inverted analytically by using the Sherman-Morrison formula: we calculate $\vect{F}_Q^{-1} = \vect{D}^{-1} - d \vect{g} \vect{g}^T$, where $d = \gamma/(1+\gamma \vect{f}^T \vect{D}^{-1} \vect{f})$ and $\vect{g} = \vect{D}^{-1}\vect{f}$.
Equation~(\ref{Eq.Sopt}) is obtained by optimizing the QCRB over the intensities of the $d$ coherent states and the probabilities $P_1, ..., P_d$ characterizing the QC operation: the minimization of $\Delta^2(\vect{v}\cdot \vect{\theta})_{\rm QCRB}$ is performed for any vector $\vect{v}$ via a chain of Cauchy-Schwarz inequalities, as detailed in \cite{supp}.
The optimal conditions are analytical and only depend on $\vect{v}$:
\be \label{Eq.optimal}
\alpha_j^{\rm (opt)} = {\rm sign}(v_j)\sqrt{ \bar{n}_c \vert v_j \vert } \quad {\rm and} \quad P_j^{\rm (opt)} = \vert v_j \vert,
\ee
regardless the state $\ket{\varphi}$.
We can further optimize the ME-L scheme over all possible choices of $\ket{\varphi}$ by maximizing $(\Delta^2 \hat{p})_{\ket{\varphi}}$ in Eq.~(\ref{Eq.Sopt}) for fixed $\bar{n}$.
The optimal state is the squeezed-vacuum~\cite{LangPRL2013}, $\ket{\varphi} = \hat{S}(r)\ket{0}$, with $\hat{S}(r) = e^{-r^2(\hat{b}^2 + (\hat{b}^\dag)^2)}$, $r > 0$ and $(\Delta \hat{p})^2\vert_{\ket{\varphi}} = e^{2r} = 2 \bar{n} + 2\sqrt{\bar{n}(\bar{n}+1)}+1$. 
Taking $\bar{n} = \bar{n}_c = \bar{n}_T/2$, we recover Eq.~(\ref{Eq.EHL})~\cite{supp}. 
This bound can be also saturated by the Schr\"odinger-cat states $\ket{\varphi} \approx (\ket{i\alpha} + \ket{-i\alpha})/\sqrt{2}$, with $(\Delta \hat{p})^2\vert_{\ket{\varphi}} \approx 4 \bar{n}$, in the limit $\bar{n} \approx \alpha^2\gg 1$.  
Optimal Fock states, $\ket{\varphi}=\ket{n}$, with $n = \bar{n}_T/2$, achieve 
\be \label{Eq.OptFock}
\min_{\ket{\varphi}=\ket{n}}\Delta^2 (\vect{v} \cdot \vect{\theta}) = \frac{2}{m[\bar{n}_T^2 + 2\bar{n}_T]}.
\ee
The additional factor 2 in Eq.~(\ref{Eq.OptFock}) with respect to Eq.~(\ref{Eq.EHL}) is related to the lack of a preferred orientation of the Fock state in the quadrature plane~\cite{PezzePRL2013, WolfNATCOMM2019}, in contrast to squeezed and Schr\"odinger cat states.

\begin{figure*}[t!]
\includegraphics[width=1\textwidth]{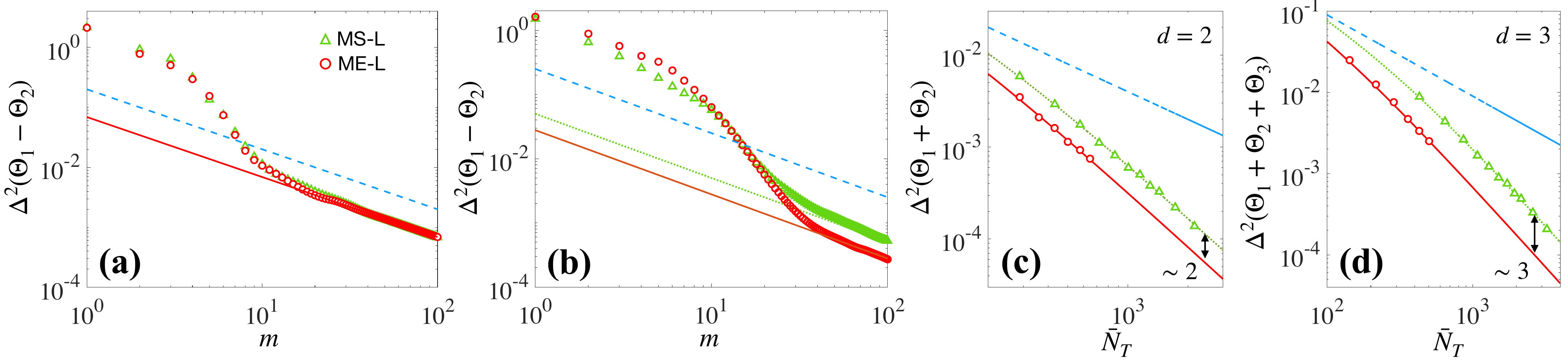}
\caption{
Maximum likelihood uncertainty for the MS-L (green triangles) and ME-L (red circles) strategies, upon imposing different constraints on resources.
(a) Comparison between MS-L and ME-L using the same probe states.
Here, the MS-L strategy uses two coherent states with $\vert \alpha_1\vert^2 = \vert \alpha_2\vert^2 = 9.5$ and two Fock states with $\ket{n}$ with $n=1$.
The ME-L strategy uses two coherent states with $\vert \alpha_1\vert^2 = \vert \alpha_2\vert^2 = 9.5$ and a single Fock state with $n=1$.
The solid red line is Eq.~(\ref{Eq.Sopt}).
(b) Comparison using the same total average number of particles per shot, $\bar{n}_T$.
Here $\bar{n}_T=16$:
The MS-L strategy uses two coherent states $\vert \alpha_1\vert^2 = \vert \alpha_2\vert^2 = 4$ and two Fock states with $n=4$.
The ME-L strategy uses two coherent states $\vert \alpha_1\vert^2 = \vert \alpha_2\vert^2 = 4$ and a single Fock state with $n=8$. 
The green dotted line and the solid red line are the QCRB, Eq.~(\ref{Eq.Sopt}), for the two strategies, respectively.
(c)-(d) Comparison using the same total average number of particles, $\bar{N}_T = \bar{n}_T \times m_{\rm opt}$, with optimized number of measurements, $m_{\rm opt} = 36$.
The two panels refer to the cases $d=2$ (c) and $d=3$ (d).
The solid red line is the fit $\sim \gamma_{\rm ME} m_{\rm opt} /\bar{N}_T^2$ for the ME-L strategies, giving $\gamma_{\rm ME} = 2.36$ and $\gamma_{\rm ME} = 2.32$ for $d=2$ and $d=3$, respectively~\cite{supp}; 
the dotted green lines is the fit $\sim \gamma_{\rm MS} m_{\rm opt} d/\bar{N}_T^2$, giving $\gamma_{\rm MS} = 2.4$.
In all panels, the blue dashed line is the SN, Eq.~(\ref{Eq.SN}). 
}
\label{Figure3}
\end{figure*}

{\it Numerics.---}We have performed a maximum-likelihood analysis of the multiparameter estimation scheme MS-L and ME-L discussed above.
For a given $\vect{\theta}$, we generate random measurement results at the output ports of the $d$ MZIs according to the conditional probability $P(\vect{\mu} \vert  \vect{\theta})$, where  $\vect{\mu} = \{n_1, m_1, ..., n_d, m_d\}$ is a vector of possible outcomes measurement of the number of particles.
We associate to a sequence of $m$ repeated independent measurements, $\vect{\mu}_m = \{ \vect{\mu}^{(1)}, ..., \vect{\mu}^{(m)} \}$, the vector of maximum likelihood estimators
\be \label{Eq.MLE}
\vect{\theta}(\vect{\mu}_m) = {\rm argmax}_{\vect{\vartheta}} \prod_{j=1}^m P(\vect{\mu}^{(j)}\vert \vect{\vartheta}).
\ee
The computation of the multidimensional probability $P(\vect{\mu} \vert  \vect{\theta})$ is restricted, for numerical convenience, to the case of Fock states $\ket{\varphi}=\ket{n}$, relatively small values of $\bar{n}_T$ and $d$.
The multiparameter maximization in Eq.~(\ref{Eq.MLE}) is performed with a gradient-descent technique.
In Fig.~\ref{Figure2}, we show various results of the maximum likelihood analysis for the ME-L scheme, in the case $d=2$. 
The different panels in  Fig.~\ref{Figure2}(a)-(d) show the two-dimensional histogram $P(\vect{\Theta} \vert \vect{\theta})$, when sensor array is optimized for a given $\vect{v}$, according to Eq.~(\ref{Eq.optimal}). 
The narrowest width of the distribution $P(\vect{\Theta} \vert \vect{\theta})$ is along the $\vect{v}\cdot \vect{\theta}$ direction, thus minimizing $\Delta^2(\vect{v} \cdot \vect{\Theta})$.
In Fig.~\ref{Figure2}(e) we consider the case $\vect{\nu} = \{0.5,-0.5\}$. 
The optimization of the sensor array guarantees that $\Delta^2 \big(\tfrac{\Theta_1 - \Theta_2}{2}\big)$ (red circles) is smaller than $\Delta^2 \big(\tfrac{\Theta_1 + \Theta_2}{2}\big)$ (blue squares) for a sufficiently large number of measurements $m$.
In particular, for $m \gg 1$, $\Delta^2 \big(\tfrac{\Theta_1 - \Theta_2}{2}\big)$ reaches Eq~(\ref{Eq.Sopt}) (solid red line), as expected, which is below the SN limit Eq.~(\ref{Eq.SN}) (dashed blue line).
This confirms numerically that the measurement of number of particles at the output ports of each MZI is indeed optimal for the saturation of the multiparameter QCRB.
In Fig.~\ref{Figure2}(f) we plot $\Delta^2 \big(\tfrac{\Theta_1 - \Theta_2}{2}\big)$ normalized to the QCRB (circles), as a function of $\vect{\theta}$ and for $m\gg1$.
The maximum-likelihood uncertainty saturates Eq~(\ref{Eq.Sopt}) (dotted lines), for all values of $\vect{\theta}$.

{\it Comparison between separable and entangled strategies.---}We first compare the MS-L and ME-L schemes when using the same probe states.
The MS-L strategy uses $d$ copies of the state $\ket{\varphi}$ as well as $d$ coherent states $\ket{\alpha_1}, ..., \ket{\alpha_d}$.
The ME-L strategy uses a single state $\ket{\varphi}$ and the coherent states $\ket{\alpha_1}, ..., \ket{\alpha_d}$.
Figure Fig.~\ref{Figure3}(a) reports the results of a numerical maximum likelihood analysis, showing $\Delta^2\big(\tfrac{\Theta_1 - \Theta_2}{2}\big)$ as a function of $m$ for the MS-L (green triangles) and ME-L (red circles) cases.
The solid red line in Fig.~\ref{Figure3}(a) shows Eq.~(\ref{Eq.Sopt}).
The numerics confirm the predicted unit gain [in the limit $m\gg 1$ and $\bar{n}_c \gg \bar{n}/(\Delta^2 
\hat{p})_{\ket{\varphi}}$] and further show that the MS-L and the ME-L schemes perform equally well in the small-$m$ regime, beyond the predictions of the QCRB.

We now compare the MS-L and the ME-L strategies with the only constraint of using the same total average number of particles per measurement shot, $\bar{n}_T$, while optimizing the distribution of resources and quantum states.
With $d$ coherent and $d$ squeezed-vacuum states as input, the MS-L reaches Eq.~(\ref{Eq.SHL}), while using $d$ coherent states and a single squeezed-vacuum the MS-L reaches Eq.~(\ref{Eq.EHL})
the highest gain is achieved for $\vect{v} = \{\pm1, \pm1, ..., \pm1\}/d$ and equals the number of parameters, $d$.
This is consistent with similar results obtained with different multiparameter estimation schemes~\cite{ProctorPRL2018, GePRL2018, GessnerPRL2018, MalitestaPRA2023}.
In our case, it is evident that this factor is due to the larger squeezing used in the ME-L scheme. 
In fact, the ME-L scheme uses a single squeezed states of $\bar{n}_T/2$ particles, while the MS-L scheme uses $\bar{n}_T/d$ particles in each MZI and $\bar{n}_T/(2d)$ particles in each of the $d$ squeezed states.
The total number of particles in the squeezed state(s) is the same in both schemes and equal to $\bar{n}_T/2$. 
However, in the ME scheme, all these resources are concentrated in a single state, while in the separable case, they are equally distributed in $d$ squeezed states. 
Similar consideration are obtained when using Fock states, $\ket{\varphi} = \ket{n}$.
In Fig.~\ref{Figure3}(b) we report the results of a maximum-likelihood analysis using optimal Fock states. 
We plot $\Delta^2\big(\tfrac{\Theta_1 - \Theta_2}{2}\big)$ as a function of $m$: we observe a gain factor $d$ (here $d=2$) for $m \gg 1$. 

Finally, we compare the MS-L and ME-L schemes for fixed total average number of particles, $\bar{N}_T = m_{\rm opt} \times \bar{n}_T$. 
Here, $m_{\rm opt}=36$ is the optimal number of measurements, evaluated separately for the MS-L and ME-L cases, respectively, for each $\bar{N}_T$~\cite{supp}.
In the numerics shown in Fig.~\ref{Figure3}(b), we have observed $m_{\rm opt} \approx 36$ in both cases, see \cite{supp} for further details. 
In Fig.~\ref{Figure3}(c) and (d) we show $\Delta^2\big(\tfrac{\Theta_1+\Theta_2}{2}\big)$ and $\Delta^2\big(\tfrac{\Theta_1+\Theta_2 + \Theta_3}{3}\big)$, for $d=2$ and $d =3$, respectively, as a function of $\bar{N}_T$.
The numerical analysis shows a gain factor $\mathcal{G} \approx 2$ and $\mathcal{G} \approx 3$ for $d=2$ and $d=3$, respectively.
It should be noticed that these results does not contradict those of Refs.~\cite{GóreckiPRL2022, GóreckiPRA2022}, where it is claimed that the advantage of ME probes does not increase with $d$, when fixing the total resources $\bar{N}_T$. 
In fact, Refs.~\cite{GóreckiPRL2022, GóreckiPRA2022} considered a figure of merit different than that the one analyzed here.
Specifically, Refs.~\cite{GóreckiPRL2022, GóreckiPRA2022} studied the sum of uncertainties $\sum_{j=1}^d \Delta^2 \theta_j$.
Instead, we use $\Delta^2(\vect{v}\cdot\vect{\theta})$, which exploits the off diagonal terms of the QFIM due to mode-entanglement in the probe state.
We have numerically verified~\cite{supp}, using the same data of Fig~\ref{Figure3}(b), that there is no gain when considering the same figure of merit as Ref.~\cite{GóreckiPRL2022, GóreckiPRA2022}.

{\it Conclusions.---}We have studied the sensitivity bounds of an array of MZIs for the estimation of an arbitrary number \textit{$d$} of phases. 
We show that the QCRB can be saturated by local measurements for every value of the phase shifts. 
We have analytically provided an optimization protocol to achieve the highest sensitivity allowed by the QCRB for arbitrary non-classical states, number of particles and the estimation of a generic combination of parameters. 
Our analysis clarifies advantages offered by combing nonclassicality of the input state and mode-entanglement created by the quantum circuit. 
We show a significant reduction of resources overhead when using the same probe states as in the  separable strategy. 
Furthermore, when fixing the total resources we show the possibility to reach a gain factor $d$, at best.
Analytical predictions are supported by a numerical maximum likelihood analysis.

\begin{acknowledgments}
We acknowledge discussions with M. Malitesta at an early stage of this work, with A. Datta and R. Demkowicz-Dobrza\`nski. 
This publication has received funding under Horizon Europe programme HORIZON-CL4-2022-QUANTUM-02-SGA via the project 101113690 (PASQuanS2.1). LP acknowledges financial support by the QuantEra project SQUEIS.
This research was supported in part by the grant NSF PHY-1748958 to the Kavli Institute for Theoretical Physics (KITP).

\end{acknowledgments}

\newpage

\begin{widetext}

\vspace{0.5cm}
\begin{center}
{\bf APPENDIX}
\end{center}

Below, we derive all equations discussed in the main text. 
For clarity, Fig.~\ref{Figure5SI}(a) and (b) we show the MS-L and ME-L schemes of Fig.~\ref{Figure1}, respectively, for the case $d=2$, clarifying the different schemes and the notation.
We note that in the appendix, we do not impose the normalization of the vector $\vect{v}$. 
Equations of the main text are recovered when imposing $\sum_{j=1}^d \vert v_j \vert =1$. 

\subsection{Quantum Fisher information matrix and multiparameter quantum Cram\'er-Rao bound}

We calculate the elements  $[\vect{F}_Q]_{j,k} = 4(\langle \Psi \vert \hat{H}_j \hat{H}_k \vert \Psi \rangle - \langle \Psi \vert \hat{H}_j \vert \Psi \rangle \langle \Psi \vert \hat{H}_k \vert \Psi \rangle)$ of the QFIM.
We recall that $\hat{H}_j = (\hat{a}_j^\dag \hat{b}_j - \hat{a}_j \hat{b}_j^\dag)/(2i)$ is the Hamiltonian of the $j$th MZI, and 
\be \label{Psi}
\ket{\Psi} = \ket{\alpha_1} \otimes ... \otimes \ket{\alpha_d} \otimes \ket{\Psi_{\rm QC}} 
=
\sum_{n_1,m_1, ..., n_d, m_d} c(n_1,m_1, ..., n_d, m_d) \ket{n_1} \ket{m_1} ... \ket{n_d} \ket{m_d},
\ee
is the overall input state of the sensor network, where $\ket{n_j}$ and $\ket{m_j}$ are fock states of the input modes $a_j$ and $b_j$ of the $j$th MZI.
In Eq.~(\ref{Psi}), the coefficients $c(n_1,m_1, ..., n_d, m_d) = c_{\rm QC}(m_1, m_2, ..., m_d) \prod_{j=1}^d (\alpha_j)^{n_j} e^{-\vert\alpha_j\vert^2/2}/\sqrt{n_j!}$ and
\be
\ket{\Psi_{\rm QC}} = \hat{U}_{\rm QC} \bigotimes_{j=1}^d \ket{\varphi_j}
=\sum_{m_1, m_2, ..., m_d} c_{\rm QC}(m_1, m_2, ..., m_d) \ket{m_1} ... \ket{m_d},
\ee
is the overall $d$-mode output state of the QC, where $\hat{U}_{\rm QC}$ is the unitary transformation representing the QC.
In our manuscript, we take real coefficients $c_{\rm QC}(n_1,m_1, ..., n_d, m_d)$ and real $\alpha_j$.
With these assumptions, we have $\bra{\alpha_j} \hat{a}_j^\dag \ket{\alpha_j} = \bra{\alpha_j} \hat{a}_j \ket{\alpha_j}$ and $\bra{\Psi_{\rm QC}} \hat{b}_j^\dag \ket{\Psi_{\rm QC}} = \bra{\Psi_{\rm QC}} \hat{b}_j \ket{\Psi_{\rm QC}}$.
Therefore 
\be
\langle \Psi \vert \hat{H}_j \vert \Psi \rangle = \frac{\bra{\alpha_j} \hat{a}_j^\dag \ket{\alpha_j} \bra{\Psi_{\rm QC}} \hat{b}_j \ket{\Psi_{\rm QC}} - \bra{\alpha_j} \hat{a}_j \ket{\alpha_j} \bra{\Psi_{\rm QC}} \hat{b}_j^\dag \ket{\Psi_{\rm QC}}}{2i} = 0, \qquad \text{for all}~j=1, ..., d,
\ee
and the QFIM becomes
\be \label{QFIMjk}
[\vect{F}_Q]_{j,k} = \alpha_j \alpha_k \bra{\Psi_{\rm QC}} \hat{b}_k^\dag \hat{b}_j + \hat{b}_j^\dag \hat{b}_k - \hat{b}_j \hat{b}_k -\hat{b}_j^\dag \hat{b}_k^\dag  \ket{\Psi_{\rm QC}} + \big( \alpha_j^2 + \bra{\Psi_{\rm QC}} \hat{b}_j^\dag \hat{b}_j \ket{\Psi_{\rm QC}} \big) \delta_{j,k}.
\ee
Computing the QFIM and its inverse requires to specify the state $\ket{\Psi_{\rm QC}}$ at the output of the QC. 
This is challenging, in general.

\begin{figure}[b!]
\includegraphics[width=0.8\columnwidth]{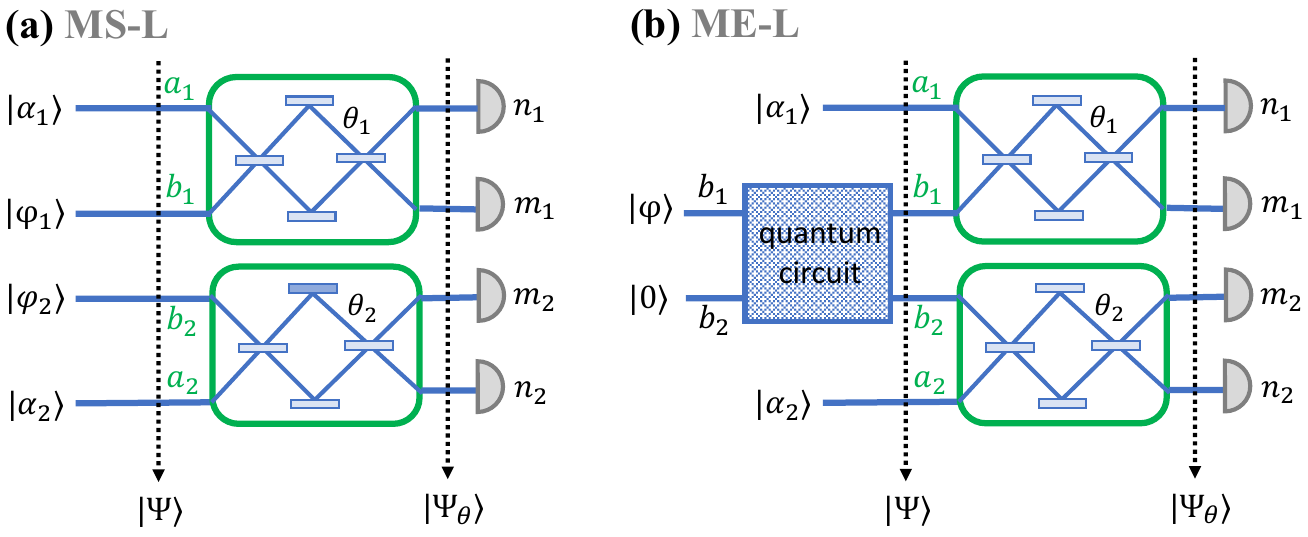}
\caption{(a) MS-L and (b) ME-L schemes in the case of $d=2$, for the estimation of relative phase shifts $\theta_1$ and $\theta_2$ in the two MZIs.
The vertical dotted line indicate the stage at which the quantum states $\ket{\Psi}$ and $\ket{\Psi_{\vect{\theta}}}$ are defined.
}
\label{Figure5SI}
\end{figure}

\subsubsection{MS-probe scheme}

The case of a MS-probe state, $\ket{\Psi_{\rm QC}} = \bigotimes_{j=1}^d \ket{\varphi_j}$, corresponds to $\hat{U}_{\rm QC} = \Eins$. 
With the assumption that $\ket{\varphi_j}$ has real coefficients when projected over Fock states, we have $\bra{\varphi_j} \hat{b}_j \ket{\varphi_j} = \bra{\varphi_j} \hat{b}_j^\dag \ket{\varphi_j}$.
Therefore, it is immediate to see that the off diagonal terms of $\vect{F}_Q$ in Eq~(\ref{QFIMjk}) vanish: $[\vect{F}_Q]_{j,k} = 0$ for $j\neq k$.
The QFIM is therefore diagonal:
\be \label{QFIM_MS_1}
[\vect{F}_Q]_{j,k} = \big[ \alpha_j^2  \bra{\varphi_j} 2 \hat{b}_j^\dag \hat{b}_j +1  - \hat{b}_j \hat{b}_j -\hat{b}_j^\dag \hat{b}_j^\dag  \ket{\varphi_j}  + \bar{n}_j \big] \delta_{j,k},
\ee
where $\bar{n}_j = \bra{\varphi_j} \hat{b}_j^\dag \hat{b}_j \ket{\varphi_j}$ is the average number of particles in the state $\ket{\varphi_j}$.
We notice that $2 \hat{b}_j^\dag \hat{b}_j + 1 - \hat{b}_j \hat{b}_j -\hat{b}_j^\dag \hat{b}_j^\dag = \hat{p}_j^2$, where $\hat{p}_j = (\hat{b}_j - \hat{b}_j^\dag)/i$ is the $p$-quadrature of mode $j$~\cite{ScullyBOOK}.
Taking $\ket{\varphi_j}$ with real coefficients also gives $\bra{\varphi_j} \hat{p}_j \ket{\varphi_j} = 0$.
Therefore, $\bra{\varphi_j} 2 \hat{b}_j^\dag \hat{b}_j +1 - \hat{b}_j \hat{b}_j -\hat{b}_j^\dag \hat{b}_j^\dag  \ket{\varphi_j} = (\Delta^2 \hat{p}_j)_{\ket{\varphi_j}}$, and we rewrite Eq.~(\ref{QFIM_MS_1}) as
\be \label{QFIM_MS_2}
[\vect{F}_Q]_{j,k} = \big[ \alpha_j^2   (\Delta^2 \hat{p}_j)_{\ket{\varphi_j}} +  \bar{n}_j  \big] \delta_{j,k}.
\ee
The QCRB for the MS-probe schemes is obtained by inverting the diagonal QFIM Eq.~(\ref{QFIM_MS_2}):
\be \label{EQ.MSQCRB}
\Delta^2(\vect{v}\cdot\vect{\theta})_{\rm QCRB} = \frac{\vect{v}^T \vect{F}_Q^{-1} \vect{v}}{m} = \frac{1}{m} \sum_{j=1}^d \frac{v_j^2}{\alpha_j^2 (\Delta^2 \hat{p}_j)_{\ket{\varphi_j}} + \bar{n}_j}.
\ee

\subsubsection{ME-probe scheme}

In the manuscript we focus in the case where the input state of the QC given by the generic state $\ket{\varphi} = \sum_{m=0}^{+\infty} c(m) \ket{m}$ in one mode and the vacuum in the other $d-1$ modes (e.g. $\ket{\varphi_j}=\ket{\varphi}$ and $\ket{\varphi_k}=\ket{0}$ for $k\neq j$).
Without loss of generality we consider the mode $b_1$ as the input mode of QC where the state $\ket{\varphi}$ is injected.
Below, we will also indicate this mode as $b$. 
Let us first show that the QC performs a multinomial splitting of the state $\ket{\varphi}$.
We have 
\begin{subequations}
\beq
\ket{\Psi_{\rm QC}} &=& \hat{U}_{\rm QC} \big[ \ket{\varphi}\otimes \ket{0} \otimes ...  \otimes \ket{0} \big] =  \\ 
&=& \sum_{m=0}^{+\infty} c(m) \, \hat{U}_{\rm QC}^\dag \big[ \ket{m}\otimes \ket{0} \otimes ...  \otimes \ket{0} \big]   \\
&=&
\sum_{m=0}^{+\infty} c(m) \, \frac{\hat{U}_{\rm QC}^\dag (\hat{b}_1^\dag)^m}{\sqrt{m!}} \big[ \ket{0}\otimes  \ket{0} \otimes ...  \otimes \ket{0} \big]  \\
&=& \sum_{m=0}^{+\infty} c(m) \, \frac{\hat{U}_{\rm QC}^\dag (\hat{b}_1^\dag)^m \hat{U}_{\rm QC}}{\sqrt{m!}} \big[ \ket{0}\otimes  \ket{0} \otimes ...  \otimes \ket{0} \big]  \\
&=& \sum_{m=0}^{+\infty} c(m) \, \frac{\big( \hat{U}_{\rm QC}^\dag \hat{b}_1^\dag\hat{U}_{\rm QC} \big)^m}{\sqrt{m!}}  \big[ \ket{0}\otimes  \ket{0} \otimes ...  \otimes \ket{0} \big], 
\eeq
\end{subequations}
where we have used, respectively, $\ket{m} = \tfrac{(\hat{b}^\dag)^m \ket{0}}{\sqrt{m!}}$, $\hat{U}_{\rm QC} \big[ \ket{0}\otimes  \ket{0} \otimes ...  \otimes \ket{0} \big] = \ket{0}\otimes  \ket{0} \otimes ...  \otimes \ket{0}$, and the unitary properties of $\hat{U}_{\rm QC}\hat{U}_{\rm QC}^\dag = \Eins$.  
The QC performs a linear transformation of the $d$ modes: $\hat{U}_{\rm QC}^\dag \hat{b}_j^\dag \hat{U}_{\rm QC} = \sum_{k=1}^d (U_{\rm QC})_{j,k} \hat{b}_k^{\dag}$, where $U_{\rm QC}$ is a $d \times d$ unitary matrix. 
Restricting to mode transformations with real coefficients, we write 
\be
\hat{U}_{\rm QC}^\dag \hat{b}_1^\dag \hat{U}_{\rm QC} = \sqrt{P_1} \hat{b}_1^\dag + \sqrt{P_2} \hat{b}_2^\dag + ... + \sqrt{P_d} \hat{b}_d^\dag,
\ee
where $P_j$ are real numbers with $\sum_{j=1}^d P_j = 1$ due to the conservation of atom number. 
The commutativity of different $\hat{b}_j$ allows to apply the multinomial theorem:
\be
\big( \hat{U}_{\rm QC}^\dag \hat{b}_1^\dag \hat{U}_{\rm QC} \big)^m = 
\big( \sqrt{P_1} \hat{b}_1^\dag + \sqrt{P_2} \hat{b}_2^\dag + ... + \sqrt{P_d} \hat{b}_d^\dag \big)^m = 
\sum_{\substack{m_1, ..., m_d =0 \\ m_1+ ... + m_d = m}}^{m} 
\frac{m!}{m_1! ... m_d!} \prod_{j=1}^m \big(\sqrt{P_j} \hat{b}_j^\dag \big)^{m_j}.
\ee
When applied to the vacuum, taking into account that $(\hat{b}_j)^{m_j}\ket{0} = \sqrt{m_j!} \ket{m_j}$, we have 
\be
\frac{\big( \hat{U}_{\rm QC}^\dag \hat{b}_1^\dag \hat{U}_{\rm QC} \big)^m}{\sqrt{m!}} \big[ \ket{0}\otimes  \ket{0} \otimes ...  \otimes \ket{0} \big] = \sum_{\substack{m_1, ..., m_d =0 \\ m_1+ ... + m_d = m}}^{m} 
\sqrt{\frac{m!}{m_1! ... m_d!}} P_1^{m_1/2} ... P_d^{m_d/2} \ket{m_1} ... \ket{m_d}. 
\ee
Finally, taking into account the sum over coefficients $c(m)$, we find that the output state of the QC is
\be \label{EQ.PsiQC}
\ket{\Psi_{\rm QC}} = \hat{U}_{\rm QC} \big[ \ket{\varphi}\otimes \ket{0} \otimes ...  \otimes \ket{0} \big] = \sum_{m=0}^{+\infty} c(m) \sum_{\substack{m_1, ..., m_d =0 \\ m_1+ ... + m_d = m}}^{m} 
\sqrt{\frac{m!}{m_1! ... m_d!}} P_1^{m_1/2} ... P_d^{m_d/2} \ket{m_1} ... \ket{m_d},
\ee 
giving
\be \label{cQC}
c_{\rm QC}(m_1, ..., m_d) = \sum_{m=0}^{+\infty} c(m)
\sqrt{\frac{m!}{m_1! ... m_d!}} P_1^{m_1/2} ... P_d^{m_d/2} \, \delta_{m_1+...+m_d,m},
\ee
where 
\be \label{Eq.Pj}
P_j = \frac{\bra{\Psi_{\rm QC}} \hat{b}_j^\dag \hat{b}_j \ket{\Psi_{\rm QC}}}{\sum_{j=1}^d \bra{\Psi_{\rm QC}} \hat{b}_j^\dag \hat{b}_j \ket{\Psi_{\rm QC}}},
\ee
is the probability to find a particle at the $j$th output mode of the QC.
We emphasize that the multinomial form of Eq.~(\ref{EQ.PsiQC}) holds since only one of the input modes of the QC is populated. 
If more than one input mode is not empty, then the form of $\ket{\Psi_{\rm QC}}$ is generally more involved.  

Using the properties of the multinomial distribution, we calculate 
\begin{subequations}
\beq
&& \bra{\Psi_{\rm QC}} \hat{b}_j^\dag \hat{b}_j \ket{\Psi_{\rm QC}} = P_j \bar{n}, \\
&& \bra{\Psi_{\rm QC}} \hat{b}_j^\dag \hat{b}_k \ket{\Psi_{\rm QC}} = \bra{\Psi_{\rm QC}} \hat{b}_k^\dag \hat{b}_j \ket{\Psi_{\rm QC}} = \sqrt{P_j} \sqrt{P_k} \bar{n}, \\
&& \bra{\Psi_{\rm QC}} \hat{b}_j \hat{b}_k \ket{\Psi_{\rm QC}} = \bra{\Psi_{\rm QC}} \hat{b}_k^\dag \hat{b}_j^\dag \ket{\Psi_{\rm QC}} = \sqrt{P_j} \sqrt{P_k} \sum_{m=0}^{+\infty} c(m)c(m+2)\sqrt{(m+2)(m+1)}, \label{bb1}
\eeq
\end{subequations}
where $\bar{n} = \bra{\varphi} \hat{b}^\dag \hat{b} \ket{\varphi}$ is the average number of particles in the state $\ket{\varphi}$.
We notice that the sum in Eq.~(\ref{bb1}) is 
$\sum_{m=0}^{+\infty} c(m)c(m+2)\sqrt{(m+2)(m+1)} = \bra{\varphi} \hat{b} \hat{b} \ket{\varphi} = \bra{\varphi} \hat{b}^\dag \hat{b}^\dag \ket{\varphi}$.
Therefore,
\be \label{bb2}
\bra{\Psi_{\rm QC}} \hat{b}_k^\dag \hat{b}_j + \hat{b}_j^\dag \hat{b}_k - \hat{b}_j \hat{b}_k -\hat{b}_j^\dag \hat{b}_k^\dag  \ket{\Psi_{\rm QC}} = \sqrt{P_j} \sqrt{P_k} \bra{\varphi} 2 \hat{b}^\dag \hat{b} - \hat{b} \hat{b} - \hat{b}^\dag \hat{b}^\dag \ket{\varphi} = (\Delta^2 \hat{p})_{\ket{\varphi}} - 1,
\ee
where $\hat{p} = (\hat{b}-\hat{b}^\dag)/i$ and the real coefficients $c(m)$ imply $\bra{\varphi} \hat{p} \ket{\varphi}=0$.
Finally, using Eqs.~(\ref{QFIMjk}) and~(\ref{bb2}) gives
\be \label{Eq.FQjk}
[\vect{F_Q}]_{j,k} = \big( \alpha_j^2 + p_j \bar{n} \big)\,\delta_{j,k} +  \big[ (\Delta^2 \hat{p})_{\ket{\varphi}}-1 \big] \big(\alpha_j \sqrt{P_j}\big) \big(\alpha_k \sqrt{P_k}\big).
\ee
We recover the QFIM in the compact form given in the main text:
\be \label{QFIM}
\vect{F}_Q = \gamma \vect{f} \vect{f}^T + \vect{D},
\ee
with $\gamma = (\Delta^2 \hat{p})_{\ket{\varphi}}-1$, $f_j = \alpha_j \sqrt{P_j}$, and $\vect{D}$ is a diagonal $d\times d$ matrix with elements $D_{jj} = \vert \alpha_j\vert^2 + P_j \bar{n}$.
The matrix Eq.~(\ref{QFIM}) can be inverted analytically thanks to the Sherman-Morrison formula:
\be
\vect{F}_Q^{-1} = \vect{D}^{-1} - \frac{\gamma}{1+\gamma \vect{f}^T \vect{D}^{-1} \vect{f}} \vect{g} \vect{g}^T,
\ee
where $\vect{g}=\vect{D}^{-1} \vect{f}$ and 
\be
\kappa = \vect{f}^T \vect{D}^{-1} \vect{f} = \sum_{j=1}^d \frac{\alpha_j^2 P_j}{\alpha_j^2 + P_j \bar{n}}.
\ee
We thus find 
\be \label{QFIMfull}
[\vect{F_Q}^{-1}]_{j,k} = \frac{\delta_{j,k}}{\alpha_j^2 + P_j \bar{n}} -
\frac{(\Delta^2 \hat{p})_{\ket{\varphi}}-1}{1+ [(\Delta^2 \hat{p})_{\ket{\varphi}}-1] \kappa} \frac{\alpha_j \sqrt{P_j}}{\alpha_j^2 + P_j \bar{n}} \frac{\alpha_k \sqrt{P_k}}{\alpha_k^2 + P_k \bar{n}}.
\ee 
Finally, the QCRB for the ME-probe schemes thus reads
\be \label{Eq.QCRB}
\Delta^2 (\vect{v} \cdot \vect{\theta})_{\rm QCRB} = \frac{\vect{v}^T \vect{F}_Q^{-1} \vect{v}}{m} = \frac{1}{m} \Bigg[ 
\sum_{j=1}^d \frac{v_j^2}{\alpha_j^2 + P_j \bar{n}} - \frac{(\Delta^2 \hat{p})_{\ket{\varphi}}-1}{1+ [(\Delta^2 \hat{p})_{\ket{\varphi}}-1] \kappa}
\Bigg( \sum_{j=1}^d \frac{v_j \alpha_j \sqrt{P_j}}{\alpha_j^2 + P_j \bar{n}} \Bigg)^2 \Bigg].
\ee

\subsection{Saturation of the Quantum Cramer-Rao bound}

For a distributed sensing scheme with unitary phase-encoding transformations $e^{-i\theta_j \hat{H}_j}$ and commuting Hamiltonians $[\hat{H}_j,\hat{H}_k] =0$ for all $j,k=1,...,d$,  Ref.~\cite{PezzePRL2017} reported necessary and sufficient conditions for which the Fisher information matrix computed with a set of projectors $\{\ket{\vect{\mu}}\bra{\vect{\mu}}\}_{\vect{\mu}}$ is equal to the QFIM. 
The condition that must be satisfied for all $j=1,\dots,d$ is~\cite{PezzePRL2017}
 \begin{equation}\label{saturation_1}
    \lim_{\tilde{\vect{\theta}}\to \vect{\theta}}\frac{\textrm{Re}[\langle\Psi_{\tilde{\vect{\theta}}}|\hat{H}_j|\vect{\mu}\rangle\langle\vect{\mu}|\Psi_{\tilde{\vect{\theta}}}\rangle]}{|\langle\vect{\mu}|\Psi_{\tilde{\vect{\theta}}}\rangle|}=0, \qquad \text{for} \,\, \langle\vect{\mu}|\Psi_{\vect{\theta}}\rangle=0,
\end{equation}
and 
\begin{equation}\label{saturation_2}
    \textrm{Re}[\langle\Psi_{\vect{\theta}}|\hat{H}_j|\vect{\mu}\rangle\langle\vect{\mu}|\psi_{\vect{\theta}}\rangle]=|\langle \Psi_{\vect{\theta}}|\vect{\mu}\rangle|^2\langle\Psi_{\vect{\theta}}|\hat{H}_j|\Psi_{\vect{\theta}}\rangle, \qquad \text{for} \,\, \langle\vect{\mu}|\Psi_{\vect{\theta}}\rangle \neq 0.
\end{equation}
A sufficient condition to fulfill both Eq.~(\ref{saturation_1}) and Eq.~(\ref{saturation_2}) is given 
\be \label{saturation_condition_main}
\textrm{Re}[\langle\Psi_{\vect{\theta}}|\hat{H}_j 
\ket{\vect{\mu}} \bra{\vect{\mu}}  \Psi_{\vect{\theta}}\rangle]=0, \qquad 
\text{for all} \, \ket{\vect{\mu}}, \, \vect{\theta} \, \text{and} \, j=1, ...d.
\ee
In fact, if this condition is satisfied, then the numerator in Eq.~\makeref{saturation_1} as well as the left-hand side of Eq.~\makeref{saturation_2}  will identically vanish. 
The right-hand side of Eq.~\makeref{saturation_2} will also vanish as a consequence of the completeness relation  $\sum_{\vect{\mu}} |\vect{\mu}\rangle\langle\vect{\mu}| = 1$ and the fact that $\hat{H}_j$ is Hermitian: $\langle\Psi_{\vect{\theta}}|\hat{H}_j|\psi_{\vect{\theta}}\rangle=\textrm{Re}[\langle\psi_{\vect{\theta}}|\hat{H}_j|\psi_{\vect{\theta}}\rangle]=\sum_{\vect{\mu}} \textrm{Re}[\langle\Psi_{\vect{\theta}}|\hat{H}_j|\vect{\mu}\rangle\langle \vect{\mu}|\Psi_{\vect{\theta}}\rangle]=0$. 
As relevant in our case (see below), Eq.~(\ref{saturation_condition_main}) can be fulfilled if $\bra{\vect{\mu}} \Psi_{\vect{\theta}} \rangle$ is real and $\bra{\vect{\mu}} \hat{H}_j \vert \Psi_{\vect{\theta}}  \rangle$ is instead imaginary (for all $j$, $\ket{\vect{\mu}}\bra{\vect{\mu}}$ and $\vect{\theta}$).

In our case, we take $\ket{\vect{\mu}} = \ket{n_1}\otimes \ket{m_1} \otimes ... \otimes \ket{n_d} \otimes \ket{m_d}$, where $\ket{n_j} \otimes \ket{m_j}$ are Fock states counting the number of particles at the $a_j$ and $b_j$ output ports, respectively, of the $j$th MZI.
The output state of the MZI array (namely before the final detection) is
\be \label{Psith}
\ket{\Psi_{\vect{\theta}}} = e^{-i \hat{\vect{H}} \cdot \vect{\theta}} \ket{\Psi} = 
\sum_{\substack{n_1,m_1, ..., n_d, m_d \\ n_1',m_1', ..., n_d', m_d'}}  c(n_1',m_1', ..., n_d', m_d') 
\mathcal{D}_{n_1',m_1'}^{n_1,m_1}(\theta_1) ...
\mathcal{D}_{n_d',m_d'}^{n_d,m_d}(\theta_d) \ket{n_1} \otimes \ket{m_1} ... \ket{n_d} \otimes \ket{m_d},
\ee
where we have used Eq.~(\ref{Psi}) and the matrix elements $\mathcal{D}_{n_j',m_j'}^{n_j,m_j}(\theta_j) = \bra{n_j} \otimes \bra{m_j} e^{-i \theta_j \hat{H}_j} \ket{n_j'}\otimes \ket{m_j'}$ are real for every $n_j$, $m_j$, $n_j'$ and $m_j'$, and every $\theta_j$~\cite{BiederharnBOOK}.
Furthermore we have 
\be \label{HPsith}
\hat{H}_j\ket{\Psi_{\vect{\theta}}} = \sum_{\substack{n_1,m_1, ..., n_d, m_d \\ n_1',m_1', ..., n_d', m_d'}} c(n_1',m_1', ..., n_d', m_d') 
\mathcal{D}_{n_1',m_1'}^{n_1,m_1}(\theta_1) ...
\mathcal{B}_{n_j',m_j'}^{n_j,m_j}(\theta_j) ...
\mathcal{D}_{n_d',m_d'}^{n_d,m_d}(\theta_d)  \ket{n_1} \otimes \ket{m_1} ... \ket{n_d} \otimes \ket{m_d},
\ee
where 
\be \label{Eq.Dtilde}
\mathcal{B}_{n_j',m_j'}^{n_j,m_j}(\theta_j) = 
\bra{n_j,m_j} e^{-i \theta_j \hat{H}_j} \hat{H}_j\ket{n_j',m_j'} = \frac{1}{2i} 
\bigg( 
\sqrt{n_j'+1} \sqrt{m_j'}
\mathcal{D}_{n_j'+1,m_j'-1}^{n_j,m_j}(\theta_j) -
\sqrt{m_j'+1}\bra{n_j,m_j}
\mathcal{D}_{n_j'-1,m_j'+1}^{n_j,m_j}(\theta_j)
\bigg).
\ee
the two terms on the right-hand side of Eq.~(\ref{Eq.Dtilde}) are real:  $\mathcal{B}_{n_j',m_j'}^{n_j,m_j}(\theta_j)$ is therefore purely imaginary.
If the coefficients $c(n_1',m_1', ..., n_d', m_d')$ in Eqs.~(\ref{Psith}) and (\ref{HPsith}) are real for all $n_j', m_j'$, then, $\bra{\vect{\mu}} \Psi_{\vect{\theta}} \rangle$ is real and $\bra{\vect{\mu}} \hat{H}_j \vert \Psi_{\vect{\theta}} \rangle$ is imaginary for all $\vect{\mu}$: Eq.~(\ref{saturation_condition_main}) is therefore fulfilled for all $j=1, ..., d$ and all $\vect{\theta}$.
This demonstrates that number counting is an optimal local measurement and the QCRB can be saturated for a sufficiently large number of measurements $m$. 

\subsection{Optimization of the QCRB for the MS schemes}

We first optimize Eq.~(\ref{EQ.MSQCRB}) over the intensities $\alpha_j^2$ ($j=1, ..., d$) of the $d$ coherent states, taking $\ket{\varphi_j} = \ket{\varphi}$ for all $j$.
We thus have $(\Delta^2 \hat{p}_j)_{\ket{\varphi_j}} = (\Delta^2 \hat{p})_{\ket{\varphi}}$ and $\bar{n}_j = \bar{n}/d$ (we recall that $\bar{n} = \sum_{j=1}^d \bar{n}_j$), giving
\be \label{LM0}
\Delta^2(\vect{v}\cdot\vect{\theta})_{\rm QCRB} = \frac{1}{m} \sum_{j=1}^d \frac{v_j^2}{\alpha_j^2 (\Delta^2 \hat{p})_{\ket{\varphi}} + \bar{n}/d}.
\ee
Imposing a constant $\bar{n}_c = \sum_j \alpha_j^2$, we write the Langrangian
\be
\mathcal{L} = \sum_{j=1}^d \frac{v_j^2}{\alpha_j^2   (\Delta^2 \hat{p})_{\ket{\varphi}} + \bar{n}/d} + \lambda \sum_{j=1}^d \alpha_j^2,
\ee
and impose $\partial \mathcal{L}/\partial \alpha_j^2=0$.
We obtain 
\be \label{LM1}
\lambda = \frac{ v_j^2 (\Delta^2 \hat{p})_{\ket{\varphi}} }{\big[\alpha_j^2   (\Delta^2 \hat{p})_{\ket{\varphi}} + \bar{n}/d\big]^2}.
\ee
We now take the square root of both terms in Eq.~(\ref{LM1}), multiply by $[\alpha_j^2   (\Delta^2 \hat{p})_{\ket{\varphi}} + \bar{n}]$ and sum over $j$.
We obtain 
\be \label{LM2}
\sqrt{\lambda} = \frac{(\sum_{j=1}^d \vert v_j \vert) \sqrt{(\Delta^2 \hat{p})_{\ket{\varphi}}}}{\big[ \bar{n}_c (\Delta^2 \hat{p})_{\ket{\varphi}} +  \bar{n}\big]}.
\ee
Considering again Eq.~(\ref{LM1}) we have 
\be \label{LM3}
\sum_{j=1}^d \frac{v_j^2}{\alpha_j^2 (\Delta^2 \hat{p})_{\ket{\varphi}} + \bar{n}/d} = \frac{\lambda}{(\Delta^2 \hat{p})_{\ket{\varphi}}} \big[ \bar{n}_c (\Delta^2 \hat{p})_{\ket{\varphi}} +  \bar{n} \big].
\ee
Finally, by combining Eqs.~(\ref{LM0}),~(\ref{LM2}) and (\ref{LM3}), we obtain 
\be
\min_{\alpha_1^2, ...\alpha_d^2} \Delta^2(\vect{v}\cdot\vect{\theta})_{\rm QCRB} = \frac{\lVert \vect{v} \rVert_1^2}{m \big[ \bar{n}_c (\Delta^2 \hat{p})_{\ket{\varphi}} + \bar{n}\big]},
\ee
where $\lVert \vect{v} \rVert_1^2 = (\sum_{j=1}^d \vert v_j \vert)^2$.
We thus recover Eq.~(\ref{Eq.Sopt}).
The optimal value of $\alpha_j^2$ is
\be
\big(\alpha_j^{(\rm opt)}\big)^2 = \frac{1}{(\Delta^2 \hat{p})_{\ket{\psi}}}\Bigg[ \frac{\vert v_j \vert}{\sum_{j=1}^d \vert v_j \vert} \big( \bar{n}_c (\Delta^2 \hat{p})_{\ket{\psi}}+\bar{n} \big) - \frac{\bar{n}}{d} \Bigg]. 
\ee

The shot noise, Eq.~(\ref{Eq.SN}), is obtained by taking $(\Delta^2 \hat{p}_j)_{\ket{\varphi}} = 1$, as in the case of the vacuum, $\ket{\varphi}=\ket{0}$, or a generic coherent state, $\ket{\varphi}=\ket{\alpha}$, where $\bar{n}_T = \bar{n}_c + \bar{n}$. \\

We now let the states $\ket{\varphi_1}, ... \ket{\varphi_d}$ be different in general, and optimize Eq.~(\ref{EQ.MSQCRB}) over all possible choices of $\ket{\varphi_j}$.
We first maximize the denominator of each terms of the sum in Eq.~(\ref{EQ.MSQCRB}).
Specifically, we maximize $(\Delta^2 \hat{p}_j)_{\ket{\varphi_j}}$ over $\ket{\varphi_j}$ for given $\bar{n}_j$.
Following Ref.~\cite{LangPRL2013}, we have that the optimal state is the squeezed-vacuum with $(\Delta^2 \hat{p}_j)_{\ket{\varphi_j}} = 2 \bar{n}_j + 2\sqrt{\bar{n}_j(\bar{n}_j+1)}+1$.
To summarise,
\be \label{LM4}
\max_{\ket{\varphi_j}} \big[ \alpha_j^2 (\Delta^2 \hat{p}_j)_{\ket{\varphi_j}} + \bar{n}_j \big] =
2 \alpha_j^2 \big[ \bar{n}_j + \sqrt{\bar{n}_j(\bar{n}_j+1)}] + \alpha_j^2 + \bar{n}_j \approx 4 \bar{n}_j \alpha_j^2 + \alpha_j^2 + \bar{n}_j,
\ee
where the last equality holds for $\bar{n}_j \gg 1$.
We further maximize Eq.~(\ref{LM4}) considering an optimal distribution of resources between the squeezed-vacuum state and the coherent state, fore fixed $\bar{n}_{T,j} = \alpha_j^2 + \bar{n}_j$.
We take the derivative of Eq.~(\ref{LM4}) with respect to $\bar{n}_j$ and use $\alpha_j^2 = \bar{n}_{T,j} - \bar{n}_j$.
We obtain $\bar{n}_j = \alpha_j^2 = \bar{n}_{T,j}/2$.
The right-hand side of Eq.~(\ref{LM4}) becomes $\bar{n}_{T,j}^2 + \bar{n}_{T,j} \approx \bar{n}_{T,j}^2$, 
such that the optimized QCRB is proportional to $\sum_{j=1}^d v_j^2/\bar{n}_{T,j}^2$.
We want to further find the optimal values of $\bar{n}_{T,j}$, for a fixed $\bar{n}_T = \sum_{j=1}^d \bar{n}_{T,j}$. 
To this aim, we write the Lagrangian
\be
\mathcal{L} = \sum_{j=1}^d \frac{\vect{v}_j^2}{\bar{n}_{T,j}^2} + \lambda \sum_{j=1}^d \bar{n}_{T,j}.
\ee 
The condition $\partial \mathcal{L}/\partial \bar{n}_{T,j}=0$ gives 
\be \label{LM5}
\lambda = \frac{k v_j^2}{\bar{n}_{T,j}^{3}}.
\ee
Equation~(\ref{LM5}) can be rewritten as 
\be \label{LM7}
\bar{n}_{T,j} = (k/\lambda)^{1/3} \vert v_j\vert^{2/3}.
\ee
Summing both sides of Eq.~(\ref{LM7}) over the index $j$ gives $\bar{n}_T = \sum_{j=1}^d \bar{n}_{T,j} = (k/\lambda)^{1/3}\sum_{j=1}^d  \vert v_j\vert^{2/3}$, that is 
\be \label{LM6}
\frac{\lambda}{k} = \Bigg( \sum_{j=1}^d \frac{\vert v_j \vert^{2/3}}{\bar{n}_T} \Bigg)^3
\ee
Finally, using Eq.~(\ref{LM5}) and (\ref{LM6}), we have 
\be
\sum_{j=1}^d \frac{\vect{v}_j^2}{\bar{n}_{T,j}^2} = \frac{\lambda}{k} \bar{n}_T
= \frac{\vert \vert \vect{v} \vert \vert_{2/3}^2}{\bar{n}_T^2},
\ee
where $\lVert \vect{v} \rVert_\kappa = (\sum_{j=1}^d \vert v_j \vert^\kappa)^{1/\kappa}$.
Using Eq.~(\ref{LM7}) and (\ref{LM6}) we find the optimal
\be
\big(\alpha_j^{(\rm opt)}\big)^2 = \bar{n}_j^{\rm (opt)} = \frac{\bar{n}_{T,j}}{2} = \frac{\bar{n}_T}{2}  \frac{\vert v_j \vert^{\tfrac{2}{3}}}{\sum_{j=1}^d \vert v_j \vert^{\tfrac{2}{3}}}. 
\ee
To conclude, we have demonstrated that  
\be
\min_{\ket{\varphi_1}, ... \ket{\varphi_d}} \sum_{j=1}^d \frac{\vect{v}_j^2}{m\bar{n}_{T,j}^k} =  \frac{\vert \vert \vect{v} \vert \vert_{2/3}^2}{m\bar{n}_T^{k}},
\ee
which recovers Eq.~(\ref{Eq.SHL}).

\subsection{Optimization of the QCRB for the ME schemes}

Now we want to optimize Eq.~(\ref{Eq.QCRB}) over the coefficients $P_j$ that characterize the QC transformation $\hat{U}_{\rm QC}$ and over the intensities $\alpha_j^2$ of the $d$ coherent states.
We demonstrate Eq.~(\ref{Eq.Sopt}):
\be \label{Eq.minim}
\min_{\alpha_1^2, ..., \alpha_d^2} \min_{P_1, ..., P_d}  \Delta^2 (\vect{v} \cdot \vect{\theta})_{\rm QCRB} = \frac{\lVert \vect{v} \rVert_1^2}{m \big[ \bar{n}_c (\Delta^2 \hat{p})_{\ket{\varphi}} + \bar{n}\big]},
\ee
where $\bar{n}_c = \sum_{j=1}^d \alpha_j^2$.
Let's neglect, for the moment, the factor $m$.
We also consider the interesting case $(\Delta^2 \hat{p})_{\ket{\varphi}} \geq 1$.
The minimization in Eq.~(\ref{Eq.minim}) is achieved by the following chain of inequalities:
\begin{subequations}
\beq \label{Eq.optCS}
\Delta^2 (\vect{v} \cdot \vect{\theta})^2_{\rm QCRB} 
&=& \sum_{j=1}^d \frac{v_j^2}{\alpha_j^2 + P_j \bar{n}} - \frac{(\Delta^2 \hat{p})_{\ket{\varphi}}-1}{1+ [(\Delta^2 \hat{p})_{\ket{\varphi}}-1] \kappa}
\Bigg( \sum_{j=1}^d \frac{v_j \alpha_j \sqrt{P_j}}{\alpha_j^2 + P_j \bar{n}} \Bigg)^2 \\
&\geq& 
\sum_{j=1}^d \frac{v_j^2}{\alpha_j^2 + P_j \bar{n}} - \frac{(\Delta^2 \hat{p})_{\ket{\varphi}}-1}{1+ [(\Delta^2 \hat{p})_{\ket{\varphi}}-1] \kappa}
\Bigg( \sum_{j=1}^d \frac{\vert v_j \vert \vert \alpha_j \vert  \sqrt{P_j}}{\alpha_j^2 + P_j \bar{n}} \Bigg)^2 \label{ineq1} \\
&\geq&
\frac{1}{1+[(\Delta^2 \hat{p})_{\ket{\varphi}}-1] \kappa} \sum_{j=1}^d \frac{v_j^2}{\alpha_j^2 + P_j \bar{n}} \label{ineq2} \\
&\geq& \frac{1}{1+[(\Delta^2 \hat{p})_{\ket{\varphi}}-1] \kappa} \frac{\lVert \vect{v} \rVert_1^2}{\bar{n}_c + \bar{n}} \label{ineq3} \\
&\geq& \frac{\lVert \vect{v} \rVert_1^2}{\bar{n}_c(\Delta^2 \hat{p})_{\ket{\varphi}}  + \bar{n}}. \label{ineq4}
\eeq
\end{subequations}

The first inequality, Eq.~(\ref{ineq1}), is obtained by noticing that
Eq.~(\ref{Eq.QCRB}) is minimized by using
\be
v_j\alpha_j \leq \vert v_j\vert \, \vert \alpha_j \vert, \qquad \text{which implies} \qquad -\Bigg( \sum_{j=1}^d \frac{ v_j  \alpha_j   \sqrt{P_j}}{\alpha_j^2 + P_j \bar{n}} \Bigg)^2 \geq - \Bigg( \sum_{j=1}^d \frac{\vert v_j \vert \vert \alpha_j \vert  \sqrt{P_j}}{\alpha_j^2 + P_j \bar{n}} \Bigg)^2,
\ee
with equality if and only if ${\rm sign}(\alpha_j) = {\rm sign}(v_j)$.
Equation~(\ref{ineq1}) follows form the fact that $0\leq \kappa \leq 1$ and the  assumption that $(\Delta^2 \hat{p})_{\ket{\varphi}} \geq 1$, giving $\tfrac{(\Delta \hat{p})^2_{\ket{\varphi}}-1}{1+ [(\Delta \hat{p})^2_{\ket{\varphi}}-1] \kappa} \geq 0$.

The second inequality, Eq.~(\ref{ineq2}), is due to Cauchy-Schwartz $(\sum_{j} X_j Y_j)^2 \leq (\sum_{j} X_j^2)(\sum_{j} Y_j^2)$ with $X_j = \vert v_j \vert / \sqrt{\alpha_j^2 + P_j \bar{n}}$ and $Y_j = \vert \alpha_j \vert \sqrt{P_j} / \sqrt{\alpha_j^2 + P_j \bar{n}}$:
\be
- \Bigg( \sum_{j=1}^d \frac{\vert v_j \vert \vert \alpha_j \vert  \sqrt{P_j}}{\alpha_j^2 + P_j \bar{n}} \Bigg)^2 \geq -\Bigg( \sum_{j=1}^d \frac{v_j^2}{\alpha_j^2 + P_j \bar{n}} \Bigg)
\Bigg( \sum_{j=1}^d \frac{\alpha_j^2 P_j}{\alpha_j^2 + P_j \bar{n}} \Bigg) = - \kappa~\Bigg( \sum_{j=1}^d \frac{v_j^2}{\alpha_j^2 + P_j \bar{n}} \Bigg) .
\ee
The inequality (\ref{ineq2}) follows, again, from the assumption $(\Delta^2 \hat{p})_{\ket{\varphi}} \geq 1$, with equality if and only if $\vert \alpha_j \vert \sqrt{P_j} = \gamma_1 \vert v_j \vert$, for some coefficient $\gamma_1$ independent from $j$.

The inequality~(\ref{ineq3}) is also due to Cauchy-Schwartz with $X_j = \vert v_j \vert/ \sqrt{\alpha_j^2 + P_j \bar{n}}$ and $Y_j = \sqrt{\alpha_j^2 + P_j \bar{n}}$:
\be
\sum_{j=1}^d \frac{v_j^2}{\alpha_j^2 + P_j \bar{n}} \geq \frac{\big( \sum_{j=1}^d \vert v_j \vert\big)^2}{\sum_{j=1}^d \alpha_j^2 + P_j \bar{n}} = \frac{\lVert \vect{v} \rVert_1^2}{\bar{n}_c + \bar{n}},
\ee
with equality if and only if $\vert v_j \vert = \gamma_2 (\alpha_j^2 + P_j \bar{n})$, for some coefficient $\gamma_2$ independent from $j$. 

The final inequality, Eq.~(\ref{ineq4}), is due to 
\be \label{ineq5}
\kappa = \sum_{j=1}^d \frac{\alpha_j^2 P_j}{\alpha_j^2 +P_j\bar{n}} \leq \frac{\bar{n}_c}{\bar{n}_c + \bar{n}}.
\ee
This follows from a maximization of $\kappa$ obtained by using the method of Lagrange multipliers, with constraints $\sum_{j=1}^d P_j=1$ and $\sum_{j=1}^d \alpha_j^2=\bar{n}_c$.
To perform the maximizatuion, we write the Lagrangian
\be
\mathcal{L} = \sum_j \frac{\alpha_j^2 P_j}{\alpha_j^2  + P_j\bar{n}} + \lambda_1 \sum_j P_j +\lambda_2 \sum_j \alpha_j^2.
\ee
Taking the partial derivatives $\partial \mathcal{L}/\partial n_j=0$ and $\partial \mathcal{L}/\partial P_j=0$ gives
\be
\lambda_1 = - \frac{\alpha_j^4}{(\alpha_j^2 + P_j \bar{n})^2}, \quad \text{and} \quad
\lambda_2 = - \frac{P_j^2 \bar{n}}{(\alpha_j^2 + P_j \bar{n})^2}.
\ee
Taking the ration between $\lambda_1$ and $\lambda_2$ and defining $\lambda\equiv \sqrt{\lambda_1/\lambda_2}$, we obtain 
\be \label{LM8}
P_j \sqrt{\bar{n}} = \lambda \alpha_j^2.
\ee
Taking the sum over the index $j$ of both sides of Eq.~(\ref{LM8}) gives $\lambda =\bar{n}_c/\sqrt{\bar{n}}$.
Replacing this expression for $\lambda$ into Eq.~(\ref{LM8}), we find 
\be \label{Popt}
P_j = \frac{\alpha_j^2}{\bar{n}_c}.
\ee
The corresponding (maximum) value of $\kappa$ is $\bar{n}_c/(\bar{n}_c + \bar{n})$.
Using Eq.~(\ref{ineq5}) into Eq.~(\ref{ineq3}) gives Eq.~(\ref{ineq4}).

Let us finally analyze the conditions for the saturation of the chain of inequalities.
We replace Eq.~(\ref{Popt}) into the condition $\vert \alpha_j \vert \sqrt{P_j} = \gamma_1 \vert v_j \vert$ discussed above. 
This gives $\alpha_j^2 = \gamma_1 \vert v_j \vert \sqrt{\bar{n}_c}$, where $\gamma_1$ can be found by summing both terms of this equation over $j$: $\gamma_1 = \sqrt{\bar{n}_c}/\sum_{j=1}^d \vert v_j\vert$.
This provides
\be \label{Popt2}
(\alpha_j^{\rm (opt)} )^2 = \frac{\bar{n}_c \vert v_j \vert}{\sum_{j=1}^d \vert v_j \vert}, \quad \text{and} \quad P_j^{\rm (opt)} = \frac{\vert v_j \vert}{\sum_{j=1}^d \vert v_j \vert}.
\ee
We notice that the condition $\vert v_j \vert = \gamma_2 (\alpha_j^2 + P_j \bar{n})$ is also satisfied by using Eq.~(\ref{Popt2}) and $\gamma_2 = (\sum_{j=1}^d \vert v_j\vert)^2/(\bar{n}_c + \bar{n})$, that is independent from $j$. 
Finally, Eq.~(\ref{Eq.optimal}) is recovered from Eq.~(\ref{Popt2}) taking into account the condition ${\rm sign}(\alpha_j) = {\rm sign}(v_j)$ discussed above, and $\sum_{j=1}^d \vert v_j \vert = \vert\vert \vect{v} \vert\vert_1$.

\subsubsection{Quadrature variance and sensitivity}

We evaluate Eq.~(\ref{Eq.Sopt}) for different example of $\ket{\varphi}$:
\begin{itemize}
    
    \item {\it Vacuum}. If $\ket{\varphi}=\ket{0}$, then $(\Delta \hat{p})^2_{\ket{\varphi}}=1$ and $\bar{n}=0$. 
    In this case Eq.~(\ref{Eq.Sopt}) recovers the shot-noise limit, Eq.~(\ref{Eq.SN}), with $\bar{n}_T = \bar{n}_c$.

    \item {\it Coherent state}. If $\ket{\varphi}=\ket{\alpha}$, then $(\Delta \hat{p})^2_{\ket{\varphi}}=1$ and $\bar{n}=\alpha^2$. 
    In this case, the QC does not create mode entanglement: the output state of the QC is a product of coherent states.
    Also in this case Eq.~(\ref{Eq.Sopt}) recovers the shot-noise limit, with $\bar{n}_T = \bar{n}_c + \alpha^2$.

    \item {\it Squeezed-vacuum states}. We consider $\ket{\varphi} = \hat{S}(r)\ket{0}$, where $\hat{S}(r) = e^{-r^2(\hat{b}^2 + (\hat{b}^\dag)^2)}$ is the squeezing operator, and $r > 0$.
    In this case, the QC generates mode entanglement and the sensitivity overcomes the SN.
    We have $(\Delta \hat{p})^2_{\ket{\varphi}} = e^{2r}$ and $\bar{n} = \sinh^2 r$. 
    By expressing $e^{2r}$ in terms of $\sinh^2 r$ we find
    \be
    (\Delta \hat{p})^2_{\ket{\varphi}} = 2\bar{n}+1 + 2\sqrt{\bar{n}(\bar{n}+1)},
    \ee 
    that is $(\Delta \hat{p})^2_{\ket{\varphi}} \approx 4 \bar{n}$ for $\bar{n}\gg 1$.
    We recall that the Mach-Zehnder sensor array with a single squeezed-vacuum state has been analyzed in Ref.~\cite{MalitestaPRA2023}, focusing on a method of moments for multiphase estimation. 

    \item {\it Fock state.} Fock states $\ket{\varphi}=\ket{n}$ are characterized by $(\Delta \hat{p})^2_{\ket{\varphi}} = 2n+1$ and $\bar{n}=n$.
    The quadrature variance $(\Delta \hat{p})^2_{\ket{\varphi}} = 2\bar{n}+1$ is larger than 1 for $n>0$.

    \item {\it Schr\"odinger-cat states.} We consider 
    \be
    \ket{\varphi}=\frac{\ket{i\alpha} + \ket{-i\alpha}}{\sqrt{2(1+e^{-2\alpha^2})}} = \sqrt{\frac{2}{1+e^{-2\alpha^2}}} \sum_{n=0}^{+\infty} \cos(n\pi/2) \frac{\alpha^n e^{-\alpha^2/2}}{\sqrt{n!}} \ket{n},
    \ee
    where $\ket{\pm i\alpha}$ are coherent states, $\alpha$ is a real number, and $\bra{i\alpha} - i \alpha \rangle = e^{-2 \alpha^2}$.
    Notice that this state has real coefficients when projected over Fock states. 
    We have 
    \be
    (\Delta \hat{p})^2_{\ket{\varphi}} = \frac{4 \alpha^2}{1+e^{-2 \alpha^2}}+1, \qquad \text{and} \qquad \bar{n} = \alpha^2 \frac{1-e^{-2 \alpha^2}}{1+e^{-2 \alpha^2}}.
    \ee
    For $\alpha^2 \gg 1$, we have $(\Delta \hat{p})^2_{\ket{\varphi}} \approx 4 \bar{n}$.

\end{itemize}

\subsection{Details on the numerical analysis}

The numerics shown in the main text refer to a maximum likelihood analysis. 
The key ingredient for the analysis is the conditional probability 
\be \label{probability}
P(\vect{\mu}\vert \vect{\theta}) =  \vert \bra{\vect{\mu}} \Psi_{\vect{\theta}}  \rangle \vert^2 = 
\bigg\vert \sum_{m=0}^{+\infty} c(m) \sum_{n_1',m_1', ..., n_d', m_d'}  \sqrt{\frac{m'!}{m_1'! ... m_d'!}} P_1^{m_1'/2} ... P_d^{m_d'/2} \delta_{m, m_1'+...+m'_d}
\prod_{j=1}^d \frac{\alpha_j^{n_j'} e^{-\alpha_j^2/2}}{\sqrt{n_j'!}}
\mathcal{D}_{n_1',m_1'}^{n_1,m_1}(\theta_1) ...
\mathcal{D}_{n_d',m_d'}^{n_d,m_d}(\theta_d) \bigg\vert^2
\ee
to observe a measurement occurrence $\vect{\mu} = \{n_1, m_1, ..., n_d, m_d\}$ for a fixed value of $\vect{\theta}$.
Equation (\ref{probability}) is obtained from Eqs.~(\ref{Psi}),~(\ref{cQC}) and~(\ref{Psith}).
Considering Fock states $\ket{\varphi}=\ket{n}$, corresponding to $c(m) = \delta_{n,m}$ in Eq.~(\ref{probability}) avoids an additional summation and simplifies the calculations. 
We also set the coherent state intensities and the probabilities $P_j$ equal to the optimal values given in Eq.~(\ref{Eq.optimal}), depending on $\vect{v}$.
To simulate experimental measurements, we randomly sample the probability $P(\vect{\mu}\vert \vect{\theta})$.
For each sequence of $m$ measurements, with outcome $\vect{\mu}_m = \{ \vect{\mu}^{(1)}, ..., \vect{\mu}^{(m)} \}$, we calculate the maximum likelihood estimator according to Eq.~(\ref{Eq.MLE}).

There are a few technical aspects in these numerics. 
First, we place a cutoff on the number of possible measurement events (which are, in principle infinite) in order to compute Eq.~(\ref{probability}).
The cutoff is such that $1-\sum_{\vect{\mu}} P(\vect{\mu}\vert \vect{\theta}) \approx 10^{-4}$. 
Second, the computation of the maximum according to Eq.~(\ref{Eq.MLE}) requires computing $P(\vect{\mu}\vert \vect{\varphi})$ on a $d$-dimensional grid $\vect{\varphi} \in [0, \pi]^d$.
A careful computation of the maximum-likelihood estimator requires the numerical grid to be much smaller than the typical width of the likelihood distribution $P(\vect{\mu}\vert \vect{\varphi})$ as a function of $\vect{\varphi}$, otherwise the sensitivity of the maximum-likelihood estimator is dominated by the grid size. 
One of the main difficulties is that the likelihood function  $P(\vect{\mu}\vert \vect{\varphi})$ shrinks around $\vect{\theta}$ by increasing $m$. 
To avoid the computational slowing down we search for the maximum of $P(\vect{\mu}\vert \vect{\varphi})$ using a gradient descent algorithm. 
Third, a single measurement sequence $\vect{\mu}_m = \{ \vect{\mu}^{(1)}, ..., \vect{\mu}^{(m)} \}$ provides a single maximum-likelihood estimate $\vect{\Theta}$. 
In order to obtain $P(\vect{\Theta}\vert \vect{\theta})$ and extract mean values and variances, we repeat the numerical experiment hundreds of times. 
In all the numerical results reported in the manuscript, we compute 
\be \label{MLS}
\Delta^2 (\vect{v} \cdot \vect{\Theta}) = \int d^d\vect{\Theta}\,P(\vect{\Theta}\vert \vect{\theta}) \, \big[ \vect{v} \cdot (\vect{\Theta}-\vect{\theta})]^2 - \bigg( \int d^d\vect{\Theta}\,P(\vect{\Theta}\vert \vect{\theta}) \, \big[ \vect{v} \cdot (\vect{\Theta}-\vect{\theta})\big] \bigg)^2.
\ee
corresponding to the mean squared fluctuations with respect to the true value $\vect{\theta}$.
This quantity takes into account not only the width of $P(\vect{\Theta}\vert \vect{\theta})$ but also the possible bias of the mean of the distribution, $\int d^d\vect{\Theta}\,P(\vect{\Theta}\vert \vect{\theta}) \vect{\Theta}$, from $\vect{\theta}$. 
\\

Details about the figures:
\begin{itemize}
    \item {\bf Figure~\ref{Figure2}(a)-(d).}
    Maximum likelihood distributions $P(\vect{\Theta} \vert \vect{\theta})$ obtained for different optimal settings of the multiparameter estimation for $d=2$.
    Specifically, we take $P_j^{\rm (opt)}$ and $\alpha_j^{\rm (opt)}$ according to Eq.~(\ref{Eq.optimal}), for a given $\vect{v}$.
    We use $n=6$ and $\bar{n}_c=6$ (such that $\bar{n}_T = 12$), $\vect{\theta}= \{\pi/2, \pi/2\}$ and $m=100$. 
    We have computed  500 values of $\vect{\Theta}$ for each panel and show Gaussian fits to the computed two-dimensional histograms. 
    The different panels correspond to different values of $\vect{v}$, which are reported in the figure.  
    
    \item {\bf Figure~\ref{Figure2}(e).}
    Maximum likelihood sensitivity, Eq.~(\ref{MLS}), as a function of the number of measurements (symbols).
    The ME-L scheme is optimized for the estimation of $(\theta_1 - \theta_2)/2$ (that is $\vect{\theta}\cdot \vect{v}$ with $\vect{v} = \{0.5,-0.5\}$).
    The distribution of resources is also optimized for $\bar{n}_T = 12$.
    That is, we take $n=\bar{n}_T/2=6$, $\bar{n}_c=\bar{n}_T/2=6$, with $\alpha_1^{\rm (opt)} = \alpha_2^{\rm (opt)} = \sqrt{3}$, according to Eq.~(\ref{Eq.optimal}).
    With this choice of parameters, the QCRB is 
    \be \label{SensFig2}
    \Delta^2 \bigg(\frac{\theta_1 - \theta_2}{2}\bigg)_{\rm QCRB} = \min_{\ket{\varphi}=\ket{n}}\Delta^2 (\vect{v} \cdot \vect{\theta}) = \frac{2}{m[\bar{n}_T^2 + 2\bar{n}_T]}, 
    \ee
    according to Eq.~(\ref{Eq.OptFock}), and 
    shown as solid red line in the figure.
    For the same parameters we also have 
    \be
    \Delta^2 \bigg(\frac{\theta_1 + \theta_2}{2}\bigg)_{\rm QCRB} = \frac{1}{m\bar{n}_T},
    \ee
    shown as dashed blue line.
    In this case, $\Delta \big(\tfrac{\theta_1 + \theta_2}{2}\big)$ coincides with the SN.
    For each value of $m$, we compute Eq.~(\ref{MLS}) from 500 repeated estimations.  
    
    \item {\bf Figure~\ref{Figure2}(f).}
    Maximum likelihood sensitivity, Eq.~(\ref{MLS}), as a function of the true value of the phase shifts $\theta_1$ and $\theta_2$.
    The ME-L scheme is optimized for the estimation of $(\theta_1 - \theta_2)/2$.
    Red circles are obtained numerically, for $n=6$ and $\bar{n}_c=6$ ($\bar{n}_T = 12$) and $m=100$
    the dashed lines are Eq.~(\ref{SensFig2}) [also coinciding with Eq.~(\ref{Eq.OptFock}) since the number of particles in the Fock state is optimized according to $n=\bar{n}_T/2$].
    Besides statistical fluctuations of $\Delta^2 \big(\tfrac{\Theta_1 + \Theta_2}{2}\big)$ due to the finite sample size (here 200 realizations for each point) and the finite $m$, we have an almost perfect agreement between the numerics and the expected QCRB.   

    \item {\bf Figure~\ref{Figure3}(a).}
    Maximum likelihood sensitivity $\Delta^2 \big(\tfrac{\Theta_1 - \Theta_2}{2}\big)$ for MS-L (green triangles) and ME-L (red circles) strategies.
    Here the two strategies are compared when using the same states. 
    The separable strategy uses two coherent states with $\alpha_1^2 = \alpha_2^2 = 9.5$ and two Fock states with $n=1$, such that $\bar{n}_T = 2(\alpha^2+1) = 21$.
    The entangled strategy uses two coherent states $\alpha_1^2 = \alpha_2^2 = 9.5$ and a single Fock state with $n=1$, such that $\bar{n}_T = \alpha_1^2+\alpha_2^2+1 = 20$.
    In both cases, the coherent states are optimized such that the sensitivity is give by Eq.~(\ref{Eq.Sopt}).
    The value of $\Delta^2(\vect{\theta}\cdot \vect{v})$ for the MS-L in Eq.~(\ref{Eq.Sopt}) is slightly smaller than the one for the ME-L due to the smaller value of $\bar{n}$: $\bar{n}=1$ for the ME-L strategy, $\bar{n}=2$ for the MS-L strategy.    
    This difference is smaller than the width of the sold red line in Fig.~\ref{Figure3}(d) and thus not visible. 
   
    \item {\bf Figure~\ref{Figure3}(b)}.
    Maximum likelihood sensitivity $\Delta^2 \big(\tfrac{\Theta_1 - \Theta_2}{2}\big)$ for MS-L (green triangles) and ME-L (red circles) strategies.
    The two strategies are compared for a fixed total average number of particles per shot, here $\bar{n}_T=16$. 
    In both cases, the numbert of particles in the Fock state is taken as the optimal one for the given $\bar{n}_T$.
    The ME-L strategy (red circles) has $n=8$ and $\alpha_1^2 = \alpha_2^2 = 4$, with  $\bar{n}_T=n+\alpha^2_1 + \alpha^2_2$.
    the MS-L strategy (green triangles) has $n=4$ and $\alpha^2=4$ [each shot consists of two measurements: one to estimate $\theta_1$ and the other to estimate $\theta_2$, such that $\bar{n}_T=2(n+\alpha^2)$].
    In both cases $\theta_1=\theta_2 = \pi/2$ and statistical average is performed over 500 independent realizations for each $m$.

    \item {\bf Figure~\ref{Figure3}(c)}.
    Maximum likelihood sensitivity $\Delta^2 \big(\tfrac{\Theta_1 + \Theta_2}{2}\big)$ for MS-L (green triangles) and ME-L (red circles) strategies.
    Here, the two strategies are compared by fixing the total average number of particles: $\bar{N}_T= \bar{n}_T \times m_{\rm opt}$, where $\bar{n}_T$ is the total number of particles per shot and $m_{\rm opt}$ is the optimal number of independent shots. 
    We find $m_{\rm opt} = 36$ from a numerical analysis.
    As an example, in Fig.~\ref{Figure5SI}(a), we plot the maximum likelihood sensitivity $\Delta^2\Theta$ for the case $d=1$, as a function of $m$ (black circles), while fixing $\bar{N}_T = \bar{n}_T \times m = 360$ (black dots).
    As we see, the minimum is obtained for $m_{\rm opt}=36$ ($\bar{n}_T = 10$).
    We have verified that $m=36$ is optimal also in the $d=2$ and $d=3$ cases and different values of $\bar{N}_T$.

    In Fig.~\ref{Figure5SI}(b) we plot the maximum-likelihood distribution $P(\Theta_1+\Theta_2)$ for the MS-L (green triangles) and ME-L (red circles) cases.
    Here $\vect{\theta} = \{\pi/2,\pi/2\}$ and $\bar{N}_T = 576$, $\bar{n}_T = 16$ and $m=36$.
    The number of particles in the Fock state is optimized for both cases, that is $n=4$ for the MS-L and $n=8$ for the ME-L.  
    Lines are Gaussian fits from which we extract $\Delta^2 \big(\tfrac{\Theta_1 - \Theta_2}{2}\big)$.
    The analysis is repeated for different values of $\bar{N}_T$ and the results are reported as symbols in Fig.~\ref{Figure3}(c).
    
    The lines in Fig.~\ref{Figure3}(c) are fits. 
    Inspired by Eq.~(\ref{Eq.OptFock}), for the ME-L scheme, we consider the fitting function (solid red line)
    \be \label{fit2d}
    f_{\rm ME}(\bar{N}_T) = \frac{\gamma_{\rm ME}}{\bar{n}_T^2 + 2\bar{n}_T} \frac{1}{m_{\rm opt}} = \gamma_{\rm ME} \frac{m_{\rm opt}}{\bar{N}_T^2 + 2\bar{N}_Tm_{\rm opt}} \approx \gamma_{\rm ME} \frac{m_{\rm opt}}{\bar{N}_T^2},
    \ee
    where the last equality is obtained for $\bar{n}_T \gg 2$ ($\bar{N}_T \gg 2 m_{\rm opt}$).
    The fit with $d=2$ gives $\gamma_{\rm ME} = 2.36$.
    For the MS-L scheme, we consider the fitting function (dotted green line)
    \be \label{fit1d}
    f_{\rm MS}(\bar{N}_T) =  \gamma_{\rm MS} \frac{ m_{\rm opt}d}{\bar{N}_T^2 + 2 \bar{N}_T d m_{\rm opt}} \approx \gamma_{\rm ME} \frac{m_{\rm opt}d}{\bar{N}_T^2}.
    \ee
    The fit with $d=2$ gives $\gamma_{\rm MS} = 2.4$. 
    For $\bar{N}_T \gg 1$ ($\bar{n}_T \gg 1$), we thus find 
    \be
    \mathcal{G} = \frac{\gamma_{\rm MS}d}{\gamma_{\rm ME}} \approx 2,
    \ee
    for $d=2$.
        
\begin{figure}[t!]
\includegraphics[width=0.8\columnwidth]{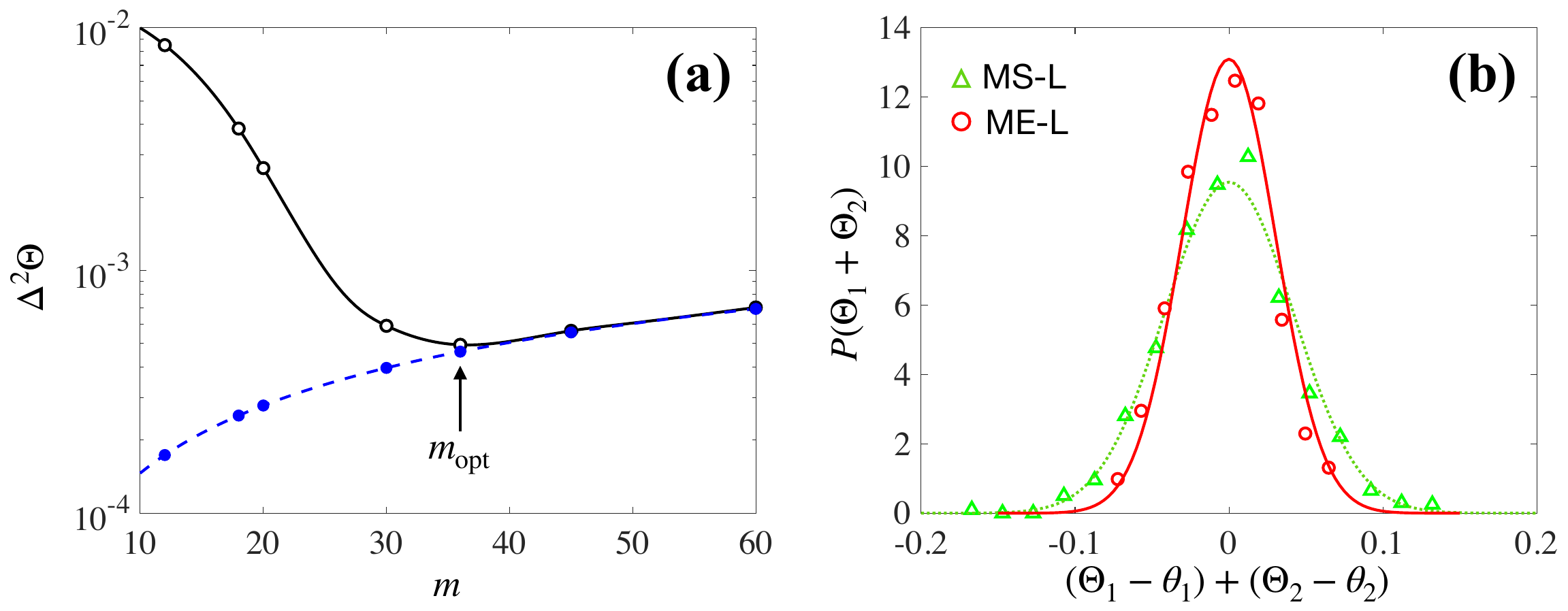}
\caption{
(a) Results of a maximum likelihood analysis for the $d=1$ case.
Here we plot $\Delta^2 \Theta$ (black circles) as a function of $m$ while fixing $\bar{N}_T = \bar{n}_T \times m = 360$.
Blue dots are the corresponding values of the QCRB.
Lines are a guide to the eye obtained by interpolating the numerical results.
(b)
Maximum likelihood distribution $P(\Theta_1+\Theta_2)$ as a function of $(\Theta_1-\theta_1)+(\Theta_2-\theta_2)$ for the MS-L (green triangles) and ME-L (red circles) strategies.
Here $\bar{n}_T = 16$ and $m=36$ in both cases.
Solid lines are Gaussian fits. 
}
\label{Figure5SI}
\end{figure}

    \item {\bf Figure~\ref{Figure3}(c).}
    Maximum likelihood sensitivity $\Delta^2\big(\tfrac{\Theta_1+\Theta_2+\Theta_3}{3}\big)$ for MS-L (green triangles) and ME-L (red circles) strategies, and fixed $\bar{N}_T$.
    Symbols are numerical results obtained with a maximum likelihood analysis.
    The lines are the fitting functions.
    The solid red line is $f_{\rm ME}(\bar{N}_T)$, given by Eq.~(\ref{fit2d}).
    The fits give $\tilde{\gamma}_{\rm ME} = 2.32$.   The dotted green line is $f_{\rm MS-L}(\bar{N}_T)$, given by Eq.~(\ref{fit1d}), with $\gamma_{\rm MS} = 2.4$.
    For $\bar{N}_T \gg 1$, we have 
    \be
    \mathcal{G} = \frac{\gamma_{\rm MS}}{\gamma_{\rm ME}d} \approx 3
    \ee
    for $d=3$.

\end{itemize}

\subsection{Sum of variances as figure of merit}

Many authors have considered, as figure of merit of multiparameter estimation, the sum of variances:
\be
V \equiv \sum_{j=1}^d \Delta^2 \theta_j \geq V_{\rm QCRB} \equiv\frac{ {\rm Tr}[\vect{F}_Q^{-1}] }{m},
\ee
where $V_{\rm QCRB}$ is the corresponding QCRB.
Using Eq.~(\ref{QFIMfull}) we find
\be
V_{\rm EQCRB} = \sum_{j=1}^d
\frac{1}{\alpha_j^2 + P_j \bar{n}} \bigg( 1 -
\frac{(\Delta \hat{p})^2_{\ket{\varphi}}-1}{1+ [(\Delta \hat{p})^2_{\ket{\varphi}}-1] \kappa} \frac{\alpha_j^2 P_j}{\alpha_j^2 + P_j \bar{n}} \bigg).
\ee
The optimization of this equation is beyond the scope of our discussion.
Here we want to compare ME-L and MS-L strategies when using the same numerical data and parameters as in Fig.~\ref{Figure3}(b) and taking $V$ as figure of merit.
The results of the analysis are reported in Fig.~\ref{Figure6}.
While in Fig.~\ref{Figure3}(b), using $\Delta^2(\Theta_1-\Theta_2)$, we observe an advantage of the entangled strategy over the separable one, when using $\Delta^2\Theta_1 + \Delta^2\Theta_2$, we observe the opposite: the separable strategy overcomes the entangled one.
The reason is that the information included in the off diagonal terms in the QFIM (and its inverse) is not entering when considering the sum of variances.
The off diagonal terms instead play an important role when taking $\Delta^2(\vect{v} \cdot \vect{\theta})$ as figure of merit, in which case the entangled strategy overcomes the separable one, as shown in the main text.

\begin{figure*}[t!]
\includegraphics[width=0.45\columnwidth]{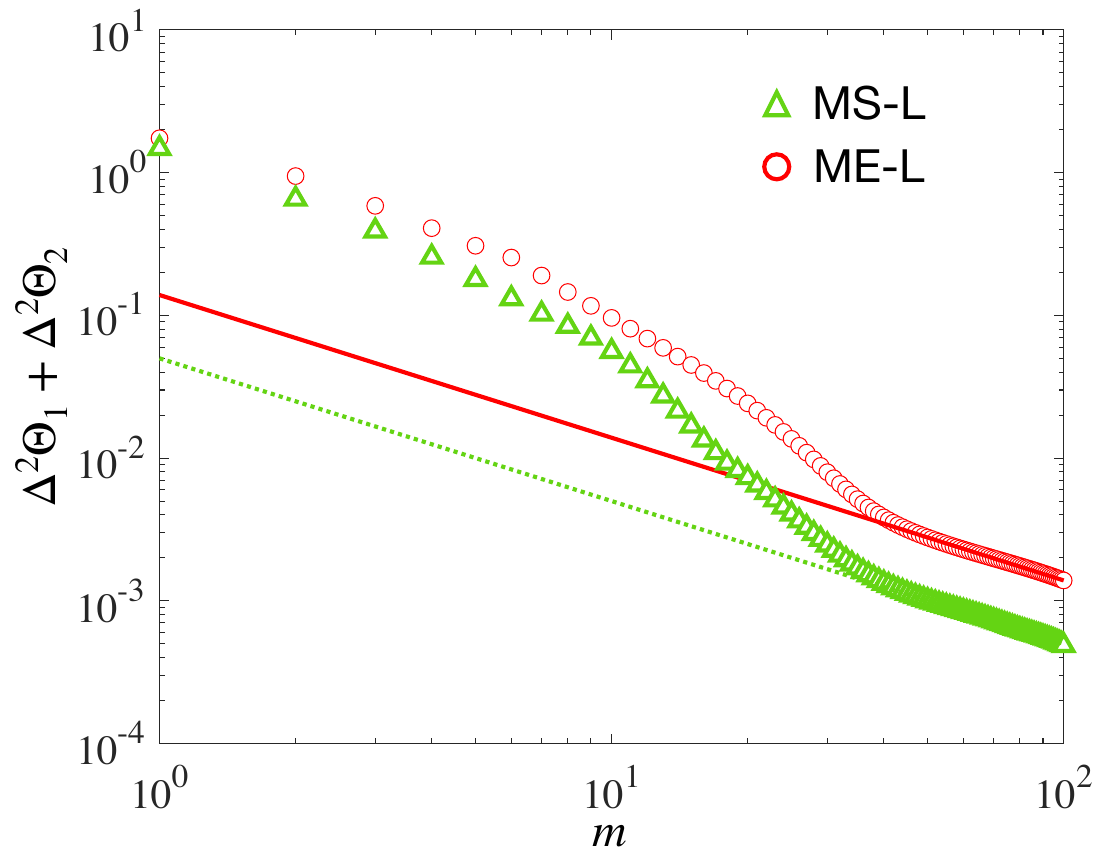}
\caption{
Comparison between ME-L (red circles) and MS-L (green triangles) strategies with the sum of variances as  figure of merit.
We plot $\Delta^2\Theta_1 + \Delta^2 \Theta_2$ obtained from a maximum likelihood analysis, as a function of $m$ (symbols)
The solid (dotted green) line is the QCRB for the ME-L (MS-S) case.
Parameters are as in Fig.~\ref{Figure3}(b). 
} \label{Figure6}
\end{figure*}

\end{widetext}


\begin{thebibliography}{100}
\bibitem{nota1}
We emphasize that we consider a genuine multiparameter problem where all $\theta_1, ..., \theta_d$ are actually estimated. 
We can thus compute any linear combination $\vect{v}\cdot \vect{\theta} = \sum_{j=1}^{d} v_j \theta_j$ of the $d$ parameters and calculate the corresponding sensitivity.
The sensor is optimized such that the estimation uncertainty is minimal for a specific $\vect{v}\cdot \vect{\theta}$, with the sensor described by the unitary evolution operator $e^{-i \sum_{j=1}^d \hat{H}_j \theta_j}$.
\end{thebibliography}
\end{document}